\documentclass[letter]{mn2e}
\usepackage{psfig}
\newcommand{\HI}{{\sc H\,i}}

\newcommand{\ltsima} {$\; \buildrel < \over \sim \;$}
\newcommand{\gtsima} {$\; \buildrel > \over \sim \;$}
\newcommand{\lta} {\lower.5ex\hbox{\ltsima}}
\newcommand{\gta} {\lower.5ex\hbox{\gtsima}}

\newcommand{\Oiii}{[{\sc O$\,$iii}]}

\newcommand{\Hb}{H$\beta$}

\newcommand{\Mgb}{Mg\,$b$}
\newcommand{\vsys}{$V_{\mathrm{sys}}$}
\newcommand{\vesc}{$V_{\mathrm{esc}}$}
\newcommand{\Mgtwee}{Mg\,2}
\title [Stellar velocity profiles and line strengths out to 4 $R_e$]
{Stellar velocity profiles and line strengths out to four effective radii in the early-type galaxies NGC\,3379 and NGC\,821}

\author[A. Weijmans et al.]
   {Anne-Marie Weijmans$^{1}$\thanks{E-mail: weijmans@strw.leidenuniv.nl}, Michele Cappellari$^{2}$, Roland Bacon$^{3}$, P.\,T. de Zeeuw$^{4,1}$, \and Eric Emsellem$^{3}$, Jes\'us Falc\'on-Barroso$^{5}$, Harald Kuntschner$^{4}$, \and
Richard M. McDermid$^{6}$,  Remco C.\,E. van den Bosch$^{7,1}$ and Glenn van de Ven$^{8}$\thanks{Hubble Fellow} \\
  $^1$ Sterrewacht Leiden, Leiden University, Postbus 9513, 2300 RA Leiden, The Netherlands \\
  $^2$ Sub-Department of Astrophysics, University of Oxford, Denys Wilkinson Building, Keble Road, Oxford OX1 3RH, UK  \\
  $^3$ Centre de Recherche Astrophysique de Lyon, University of Lyon, 9 Avenue Charles Andr\'e, 69230 Saint Genis Laval, France \\
  $^4$ European Southern Observatory, Karl-Schwarzschild-Str 2, 85748 Garching, Germany \\
 $^5$ European Space and Technology Centre, Postbus 299, 2200 AG Noordwijk, the Netherlands \\
 $^6$ Gemini Observatory, Northern Operations Centre, 670 N. A'ohoku Place, Hilo, Hawaii 96720, USA \\
 $^7$ Astronomy Department, University of Texas, Austin, TX 78712, USA \\
 $^8$ Institute for Advanced Study, Einstein Drive, Princeton, NJ 08540, USA 
   }

\pagerange{\pageref{firstpage}--\pageref{lastpage}} \pubyear{2009}

\begin{document}

\maketitle

\label{firstpage}

\begin{abstract} 
  We use the integral-field spectrograph SAURON to measure the stellar
  line-of-sight velocity distribution and absorption line strengths
  out to four effective radii ($R_e$) in the early-type galaxies
  NGC~3379 and NGC~821. With our newly developed observing technique
  we can now probe these faint regions in galaxies that were
  previously not accessible with traditional long-slit
  spectroscopy. We make optimal use of the large field-of-view and
  high throughput of the spectrograph: by adding the signal of all
  $\sim$1400 lenslets into one spectrum, we obtain sufficient
  signal-to-noise in a few hours of observing time to reliably measure
  the absorption line kinematics and line strengths out to large
  radius.

  We find that the line strength gradients previously observed within
  1 $R_e$ remain constant out to at least 4 $R_e$, which puts
  constraints on the merger histories of these galaxies. The stellar
  halo populations are old and metal-poor. By constructing orbit-based
  Schwarzschild dynamical models we find that dark matter is necessary
  to explain the observed kinematics in NGC~3379 and NGC~821, with 30
  - 50 per cent of the total matter being dark within 4 $R_e$. The
  radial anisotropy in our best-fit halo models is less than in our
  models without halo, due to differences in orbital structure. The
  halo also has an effect on the \Mgb\ - \vesc\ relation: its slope is
  steeper when a dark matter halo is added to the model.

\date{}
\end{abstract}


%%%%%%%%%%%%%%%%%%%%%%%%%%%%%%%%%%%%%%%%%%%%%%%%%%%%%%%%%%%%%%%%

\section{Introduction}
\label{sec:introduction}

According to the current galaxy formation paradigm, galaxies should be
embedded in massive dark matter haloes (e.g. Kauffmann \& van den
Bosch 2002\nocite{2002SciAm.286f..36K}; Springel et
al. 2005\nocite{2005Natur.435..629S}). For spiral galaxies, these
haloes have indeed been observationally confirmed, by determining
their influence on the kinematics of the \HI\ discs that surround most
of these galaxies (see e.g. van Albada et
al. 1985\nocite{1985ApJ...295..305V}). Early-type galaxies often lack
these large gas discs, and therefore their dark matter haloes are more
difficult to probe. Although some early-type galaxies do have regular
\HI\ structures that have been used to identify a dark matter halo
(e.g. Bertola et al. 1993\nocite{1993ApJ...416L..45B}; Franx, van
Gorkom \& de Zeeuw 1994\nocite{1994ApJ...436..642F}; Oosterloo et
al. 2002\nocite{2002AJ....123..729O}; Weijmans et
al. 2008\nocite{2008MNRAS.383.1343W}) and recent sensitive radio
observations have shown that \HI\ in these galaxies is more common
than previously thought (e.g. Morganti et al.
2006\nocite{2006MNRAS.371..157M}), for the majority of early-type
galaxies we need to employ other tracers to determine the dark matter
content.

Stellar kinematics are a reliable tracer of the gravitational
potential in galaxies. Unfortunately, traditional long-slit
spectroscopy seldom reaches beyond two half-light or effective radii
($R_e$), because the necessary signal-to-noise ratio ($S/N$) is
difficult to achieve in this regime. In the central parts of galaxies
this is not an issue, and it has been shown by various studies that in
early-type galaxies within one $R_e$ the luminous matter still
dominates over the dark matter: only 10 - 50 per cent of the total
mass is dark (Gerhard et al. 2001\nocite{2001AJ....121.1936G};
Cappellari et al. 2006\nocite{2006MNRAS.366.1126C}; Thomas et
al. 2007\nocite{2007MNRAS.382..657T}). These data however do not cover
the outer regions of galaxies, where the dark matter is supposed to
dominate. Models of early-type galaxies based on long-slit stellar
kinematics extending to larger radii are not always conclusive, but
indicate that a dark halo is required in some individual cases
(e.g. Carollo et al.  1995\nocite{1995ApJ...441L..25C}; Rix et al.
1997\nocite{1997ApJ...488..702R}).

To probe the gravitational potential at larger radii (outside one
$R_e$), we can resort to gravitational lensing (e.g. Koopmans et
al. 2006\nocite{2006ApJ...649..599K}; Gavazzi et
al. 2007\nocite{2007ApJ...667..176G}), X-ray emission (e.g. O'Sullivan
\& Ponman 2004\nocite{2004MNRAS.354..935O}; Humphrey et
al. 2006\nocite{2006ApJ...646..899H}) or discrete tracers such as
globular clusters (e.g. C\^ot\'e et
al. 2003\nocite{2003ApJ...591..850C}; Bridges et
al. 2006\nocite{2006MNRAS.373..157B}) or planetary nebulae (e.g.
Romanowsky et al. 2003\nocite{2003Sci...301.1696R}; Douglas et
al. 2007\nocite{2007ApJ...664..257D}; Napolitano et
al. 2009\nocite{2009MNRAS.393..329N}). Interestingly, Romanowsky et
al. (2003)\nocite{2003Sci...301.1696R} found that models of their
sample of three intermediate luminosity early-type galaxies only
required a small amount of dark matter, if any at all. Dekel et
al. (2005)\nocite{2005Natur.437..707D} subsequently argued that the
observed kinematics of the planetary nebulae (PNe) could also be
explained by invoking radial anisotropy, and therefore do not exclude
a dark halo (see also Binney \& Mamon
1982\nocite{1982MNRAS.200..361B}). Douglas et
al. (2007)\nocite{2007ApJ...664..257D} presented newly reduced PNe
kinematics for one of the galaxies of the Romanowksy et al. sample
(NGC~3379) and found the dark matter fraction within 5 $R_e$ to be
less than 40 per cent, which is significantly lower than what is
predicted by $\Lambda$CDM models. Napolitano et
al. (2009)\nocite{2009MNRAS.393..329N} reached similar conclusions
modeling PNe kinematics in the elliptical galaxy NGC~4494 and
suggested that intermediate-luminosity galaxies are surrounded by low
concentration haloes, while some high-mass ellipticals could have more
strongly concentrated haloes. De Lorenzi et
al. (2009)\nocite{2009MNRAS.395...76D} presented mass models of
NGC~3379 based on the PNe kinematics of Douglas et
al. (2007)\nocite{2007ApJ...664..257D}, combined with long-slit and
central SAURON integral-field data, and concluded that they can fit the
data with a range of mass distributions, depending on the
anisotropy. They did not exclude a dark matter halo with a strong
radially anisotropic outer envelope, that is still consistent with
merger models within the $\Lambda$CDM paradigm.

We observed two of the galaxies of the Romanowsky et
al. (2003)\nocite{2003Sci...301.1696R} sample, NGC~3379 and NGC~821,
with the SAURON integral-field unit (IFU), to measure the
line-of-sight velocity distribution (LOSVD) at large radii (3-4
$R_e$). We took advantage of the high throughput and large
field-of-view of SAURON (41 $\times$ 33 arcsec$^2$) and used the IFU
as a ``photon collector'': for each observation we co-added all
spectra within one field to acquire enough $S/N$ to measure the LOSVD
up to the fourth Gauss-Hermite moment $h_4$. These higher order
moments are key to breaking the mass-anisotropy degeneracy when
constructing mass models and determining the dark matter content of
galaxies (e.g. Gerhard 1993\nocite{1993MNRAS.265..213G}). Apart from
the LOSVD, we also obtained line strengths from our spectra, allowing
us to study line strength gradients out to large radii and
constraining the properties of the stellar halo populations. Both
NGC~3379 and NGC~821 are well studied galaxies, and some of their
properties are summarized in Table~\ref{tab:gal}. Their central
regions have been observed as part of the SAURON survey (de Zeeuw et
al. 2002)\nocite{2002MNRAS.329..513D}. They are classified as fast
rotators (Emsellem et al. 2007)\nocite{2007MNRAS.379..401E} and
display regular stellar kinematics.

\begin{table}
\label{tab:gal}
\begin{center}
\begin{tabular}{l|r|r}
\hline\hline
Parameter & NGC 821  & NGC 3379\\
\hline
Morphological Type & E6? & E1 \\
$M_B$ (mag) & -20.74 & -20.57 \\
Effective $B-V$ (mag) & 0.87 & 0.93 \\
Photometric PA ($^{\circ}$) & 32 & 68 \\
Distance modulus (mag) & 31.85 & 30.06 \\
Distance (Mpc) &  23.4 & 10.3 \\
Distance scale (pc arcsec$^{-1}$) &  114 & 50 \\ 
Effective radius (arcsec) & 39  & 42 \\
Systemic velocity (km/s) & 1726 & 930 \\
\hline
\end{tabular}
\end{center}
\caption{Properties of NGC~821 and NGC~3379. Magnitude and colour are taken from the Lyon/Meudon Extra galactic Database (LEDA). The distance modulus was derived from surface brightness fluctuation measurements by Tonry et al. (2001)%\nocite{2001ApJ...546..681T}
. Note that 0.06 mag is subtracted to adjust to the Cepheid zeropoint of Freedman et al. (2001)%\nocite{2001ApJ...553...47F}
; see Mei et al. (2005), %\nocite{2005ApJ...625..121M}
 section 3.3, for a discussion. The effective radius and position angle are taken from Cappellari et
 al. (2007).} %\nocite{2007MNRAS.379..418C}

\end{table}

In section 2, we describe our observations and data reduction. We
present our spectra and discuss the resulting LOSVD in section
3. Section 4 is devoted to measuring line strengths and we investigate
gradients and the properties of the stellar populations at large
radii. In section 5 we present our mass models of NGC~3379 and NGC~821
(including a dark matter halo) and connect the dynamical properties
with stellar populations by constructing the \Mgb\ - \vesc\
relation. Section 6 summarizes and discusses our results.

%%%%%%%%%%%%%%%%%%%%%%%%%%%%%%%%%%%%%%%%%%%%%%%%%%%%%%%%%%%%%%%%%

\section{Observations and data reduction}
\label{sec:observations}

We observed four fields in NGC 3379 with SAURON at the William
Herschel Telescope at La Palma, Spain, in March 2005. The fields were
centred at 2.6 and 3.5 $R_e$, on opposite sides of the nucleus to
allow for an evaluation of the systemic velocity of the system. They
were not placed on the major axis of the galaxy but instead on an axis
that deviates by 23$^{\circ}$. This was done to avoid contamination in
our outer field by the neighbouring galaxy NGC 3384. The positions of
the fields were chosen such that most of them did not contain
foreground stars, and they were oriented such that the skylenslets of
SAURON pointed away from the galaxy nucleus, obtaining skyspectra 2
arcminutes away from the SAURON field.

\begin{figure*}
\begin{tabular}{cc}
\psfig{figure=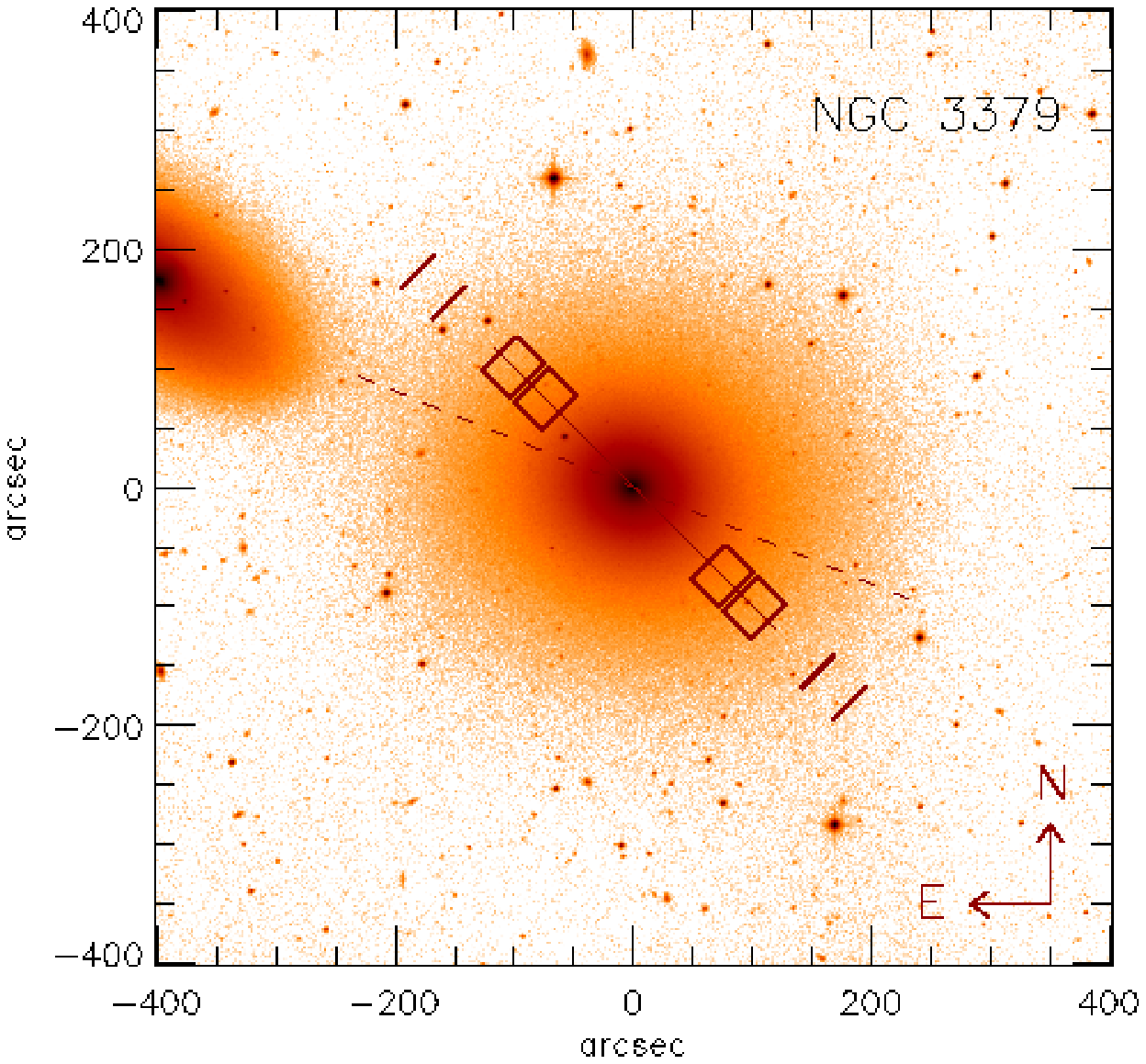,angle=0,width=7.0cm} & \psfig{figure=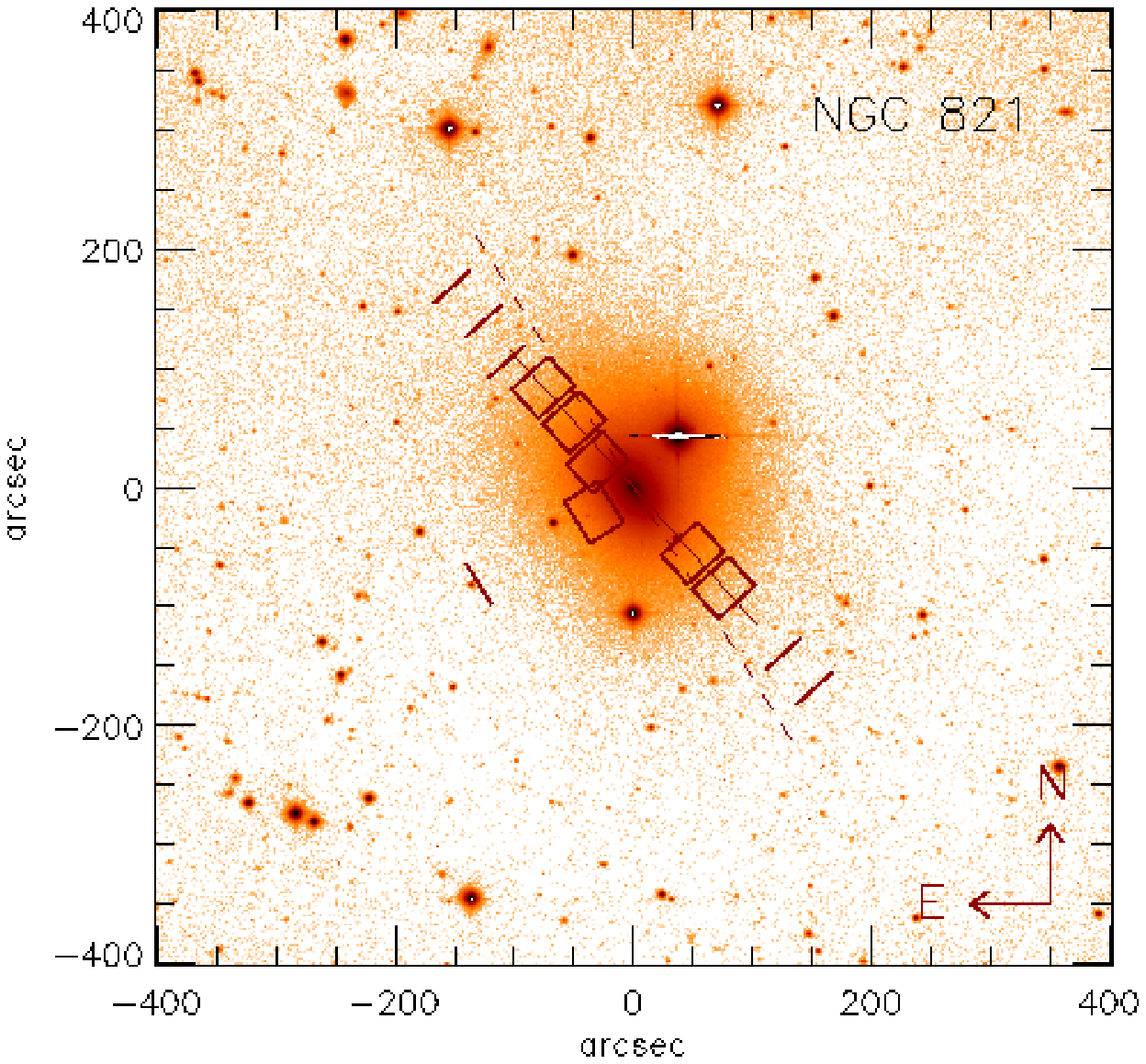,angle=0,width=7.0cm}
\end{tabular}
\caption{Positions of our observed fields in NGC 3379 (left) and NGC 821 (right). The red boxes denote each SAURON field of view. The skylenslets (red short thick lines) are aligned with the long side of the SAURON field, at a distance of two arcminutes. SAURON was orientated such that the skylenslets pointed away from the galaxy nucleus. The fields are not exactly aligned with the major axis of the galaxy (denoted by the dashed line) to minimize the contamination of foreground stars and neighbouring galaxies. The underlying images were obtained with the 1.3-m McGraw-Hill Telescope at MDM Observatory.}
\label{fig:posfields}
\end{figure*}

We used a similar strategy to observe six fields in NGC 821 in
September 2006. One field was situated on the minor axis, one field at
1 $R_e$, two fields at opposite sides of the galaxy nucleus at 2 $R_e$
and the last two fields on opposite sides of the nucleus at 3 $R_e$.
Figure~\ref{fig:posfields} shows the observed fields in NGC 3379 and
NGC 821, and Table~\ref{tab:exp} summarizes their exposure times and
exact positions.

\begin{table*}
\label{tab:exp}
\begin{center}
\begin{tabular}{l|l|c|c|c|c|c|l}
\hline\hline
Galaxy & Field & $R / R_e$  & T$_\mathrm{exp}$ & $\mu_V$   &  $S/N$ & galaxy contr. & Comments\\
 & & & (hr) &  (mag/arcsec$^2$)&  & (per cent) & \\ 

\hline
NGC 3379 & 3N & 2.6 & 2.0 & 23.0 & - & -  & cloudy conditions, no signal \\
 & 3S & 2.6 & 2.5 & 23.0 & 29 & 15 & \\
 & 4N & 3.5 & 7.5 & 23.7 & 16 & 5  & cloudy conditions \\
 & 4S & 3.5 & 2.0 & 23.7 & 21 & 13 & \\
\hline
NGC 821 & minor & 1.0 & 1.5 & 23.0 & 58 & 28 &\\
 & 1N & 0.9 & 1.5 & 21.7 & 57 & 24 &\\
 & 2N & 1.9 & 1.5 & 23.3 & 12 & 7 & cloudy conditions \\
 & 2S & 1.9 & 0.5 & 23.3 & - & - & cloudy conditions, no signal \\
 & 3N & 2.9 & 4.0 & 24.5 & - & -  & cloudy conditions, no signal \\
 & 3S & 2.9 & 4.0 & 24.5 & 9 & 4 & \\
\hline

\end{tabular}
\caption{Properties of our observed fields in NGC 3379 and NGC 821. The fields are named according to their distance and orientation with respect to the galaxy centre, see Figure~\ref{fig:posfields}. The surface brightness $\mu_V$ is calculated from the MDM images, and can be compared to the sky surface brightness, which in $V$-band is about 21.7 mag/arcsec$^2$ in optimal conditions at our observed fields (Benn \& Ellison 1998).} %\nocite{1998NewAR..42..503B
\end{center}
\end{table*}

The data were reduced using the dedicated \textsc{xsauron} software
(Bacon et al. 2001\nocite{2001MNRAS.326...23B}). The shutter
of SAURON malfunctioned during the observing run of NGC 821, and did
not close during read-out of the CCD. This did not affect our science
frames, whose exposure times of 1800 seconds are long compared to the
typical read-out time of the CCD ($\sim$ 30 seconds). However, our
calibration frames have shorter exposure times (15 - 60 seconds) and
were badly contaminated by light hitting the CCD during read-out. We
could therefore not rely on these calibration frames. Instead, we took
the calibration frames obtained during the observing run of NGC 3379
and rotated and translated them so that they coincided with the
contaminated frames that they would replace. The spectra of NGC 821
were then extracted and flat-fielded using these frames.

We checked that the extraction of the spectra was done accurately, but
we found residuals in the frames after flat-fielding. This was to be
expected: although the extraction mask is rather stable between
observing runs, the flat-field is not. We therefore constructed a
superflat by combining our six empty frames with 30 frames of a Lyman
$\alpha$ emitting halo, taken during the same run (see Weijmans et
al. in prep. for details). Since these 36 frames contain (mostly)
empty sky and are offset with respect to each other, their median,
smoothed in the spectral direction, proved to be a suitable superflat
(see also Bower et al. 2004\nocite{2004MNRAS.351...63B}). Dividing our
already flat-fielded spectra by this superflat removed most of the
residuals. Coincidently, we could also construct a superflat for our
NGC 3379 dataset: immediately after our observing run SAURON was used
in the same setup to observe Lyman $\alpha$ emission around high
redshift submillimeter galaxies. These frames also consist mostly of
empty sky, and 19 of them were used to create a superflat for NGC
3379.

After superflat-fielding, both datasets were reduced in the standard
manner, except that we did not subtract the sky from the spectra. We
will take the contribution of the sky into account when fitting the
spectra to extract the stellar kinematics (see Section~\ref{sec:losvd}).

%%%%%%%%%%%%%%%%%%%%%%%%%%%%%%%%%%%%%%%%%%%%%%%%%%%%%%%%%%%%%%%%%%

\section{Stellar kinematics}
\label{sec:losvd}

To extract the stellar kinematics from our spectra, we used the
penalized pixel fitting method (pPXF)\footnote{available from
www-astro.physics.ox.ac.uk/$\sim$mxc/idl/\#ppxf} of Cappellari \&
Emsellem (2004)\nocite{2004PASP..116..138C}. This method fits a
stellar template spectrum, convolved with a line-of-sight velocity
distribution (LOSVD), to the observed galaxy spectrum in pixel space
(i.e. logaritmically binned in wavelength). If apart from the mean
velocity $V$ and velocity dispersion $\sigma$ also the Gauss-Hermite
moments $h_3$ and $h_4$ are fitted, then the solution is biased
towards a Gaussian. The amount of biasing can be controlled, and
following the recipe described in Cappellari \& Emsellem
(2004)\nocite{2004PASP..116..138C} we found that a penalization factor
of $\lambda$ = 0.5 gives the best results for our dataset and
signal-to-noise.

\subsection{Method}

\begin{figure}
\centerline{\psfig{figure=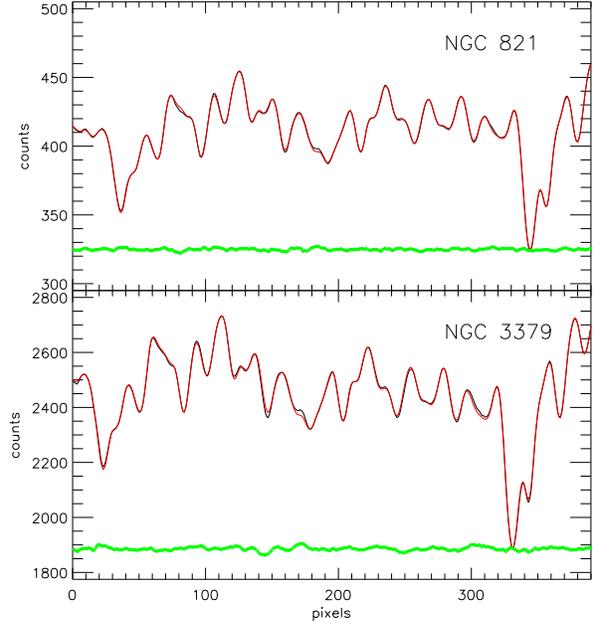,angle=0,width=8.0cm}
}
\caption{Co-added spectra of NGC 821 and NGC 3379 (black line) and
overplotted their optimal stellar templates broadened with the best-fitting
LOSVD (red line). Residuals (green, shown with offset for
presentation purposes) are small, and due to template mismatch.}
\label{fig:spec_cen}
\end{figure}

To find a suitable stellar template to fit our spectra, we co-added all
the spectra of the central SAURON fields of NGC 821 and NGC 3379,
presented in Emsellem et al. (2004)\nocite{2004MNRAS.352..721E}. We then
determined the optimal template for these high signal-to-noise spectra
as a linear superposition from the full MILES library, containing 945
stars (S\'anchez-Bl\'azquez et
al. 2006\nocite{2006MNRAS.371..703S}). For NGC 3379, a template
consisting of 19 stars was created, and for NGC 821 a template of 21
stars. Template mismatch is negligible, as can be seen in
Figure~\ref{fig:spec_cen}.  We used these templates to measure the
stellar kinematics in our spectra at large radii in NGC 3379 and NGC
821. Additive polynomials were included in our fit to compensate for
changes in line strengths, which are to be expected over the radial
range that we cover. We found that varying the stellar templates gave
similar results for the kinematics within the errors.

Since our spectra are very skydominated (see Table~\ref{tab:exp}), we
did not subtract the sky from our spectra, but instead included the
skyspectra obtained by co-adding the spectra observed with the
$\sim$170 skylenslets of SAURON (pointing $\sim$ 2 arcminutes away
from the field of view) for each exposure as separate templates into
pPXF. While the stellar template was broadened with the LOSVD, at the
same time the sky spectra were scaled to minimize the $\chi^2$. The
modeled spectrum $G_{\mathrm{mod}}$, with $x$ in pixelspace, is then
described by:

\begin{equation}
\label{eq:ppxf}
G_{\mathrm{mod}}(x) = [B \otimes T](x) + \sum_{l=0}^L b_l\mathcal{P}_l(x) + 
   \sum_{n=1}^N w_n S_n(x),
\end{equation}

\noindent
with $B(x)$ the broadening function, $T(x)$ the optimal stellar
template, $\mathcal{P}_l(x)$ the $L+1$ additive polynomials of degree
$l$ with coefficients $b_l$, and $S_n(x)$ the $N$ sky templates, with
weight $w_n \ge 0$ (see also Eq.~3 in Cappellari \& Emsellem
2004\nocite{2004PASP..116..138C}). By fitting the contribution of the
sky in this way instead of subtracting the skyspectra directly, we
allow the inclusion of subtle sky variations. We found that especially
for the spectra with a low galaxy contribution (see
Table~\ref{tab:exp}) this improved the fits significantly.

We subtracted the resulting composite skytemplate
$\sum_{n=1}^{N}w_nS_n(x)$ from the observed spectrum, and ran pPXF
again. Errors are estimated by a Monte Carlo method, where we measured the
kinematics from many realisations of the input spectrum, on which we
added noise. The noise spectrum was obtained from the difference
between our observed spectrum and the best-fit modeled spectrum
($G_{\mathrm{mod}}$). We found that varying the amount of subtracted
sky by up to 10 per cent does not affect our measured LOSVD. This
shows that our LOSVD parameters are robust to errors in our sky
subtraction. We also checked that the measured LOSVD is indeed the
LOSVD of the galaxy and not an imprint of the solar spectrum from
e.g. zodiacal light: when measuring the LOSVD of the sky templates
themselves we measured a velocity offset with respect to the galaxy
comparable to the systemic velocity of the system, and a much smaller
broadening.

We found that the best fits were obtained by including the skyspectra
from the skylenslets, taken simultaneously with our galaxy
spectra. Providing the sky spectra of our blank skyfields as templates
gave worse results, even though these fields covered a larger sky area
than the skylenslets and therefore had higher $S/N$. This indicates
the importance of obtaining simultaneous skyspectra over high $S/N$,
to avoid mismatch due to the variability of the night sky.

Finally, we tested the influence of subtracting residual galaxy light
that could be present in our sky spectra, since the skylenslets were
pointing at 6 - 7 $R_e$ in the galaxy. The galaxy light at these
distances of the nucleus is however very faint, approximately 3-4
mag/arcsec$^2$ fainter than in the regions where we measure our
kinematics. We simulated the effect of subtracting such a weak
Gaussian absorption line from our observed line and found that the
maximal error we can introduce in this way is 8 km/s in our
measured velocity dispersions. This is well within our error bars, and
we conclude that this effect is negligible.

\subsection{NGC 3379}

\begin{figure}
\centerline{\psfig{figure=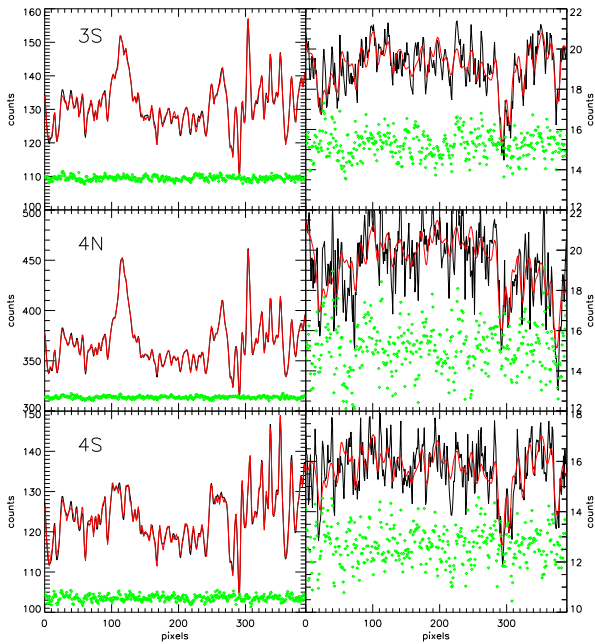,angle=0,width=8.0cm}
}
\caption{Spectra of our fields in NGC 3379. The left panels show the
observed spectra (not skysubtracted) in black, with overplotted in red
the best fit from pPXF (combination of stellar template broadened with
LOSVD and scaled sky templates). The residuals (green) are offset for
presentation purposes. Right: skysubtracted spectra (black) with the
same broadened LOSVD (red) as in the left panels. Residuals are again shown in green.}
\label{fig:ppxf_3379}
\end{figure}

It turned out that one of our fields in NGC 3379, 3N, did not have
sufficient signal to measure the LOSVD. The results for the other fields can
be found in Figure~\ref{fig:ppxf_3379} and
Table~\ref{tab:kin3379}. The last four columns for
Table~\ref{tab:kin3379} show the results for the LOSVD if we restrict
our fit to the first two moments, instead of fitting up to $h_4$. The
results for both fits agree within the errors.

 Correcting for barycentric motion, we find that the systemic velocity
\vsys\ of the galaxy measured from the central field (presented by
Emsellem et al. 2004\nocite{2004MNRAS.352..721E}) is 930 $\pm$ 2 km/s.
This agrees with \vsys\ as determined from the 3S and 3N field pair:
although we could not measure the velocity dispersion from the
spectrum in 3N, we were able to determine the positions of the
absorption lines in the spectrum and therefore the velocity offset
with respect to the 3S field, opposite the nucleus. By averaging the
two velocities from these field we find \vsys\ = 938 $\pm$ 16
km/s. However, repeating this exercise for the field pair at 3.5
$R_e$, which are 4S and 4N, we find \vsys\ = 892 $\pm$ 23 km/s, which
seems marginally deviant from our other two measurements, although
this is just outside the 1-$\sigma$ interval. We therefore decided to
subtract for each field pair their own corresponding systemic
velocity.

\begin{figure}
\centerline{\psfig{figure=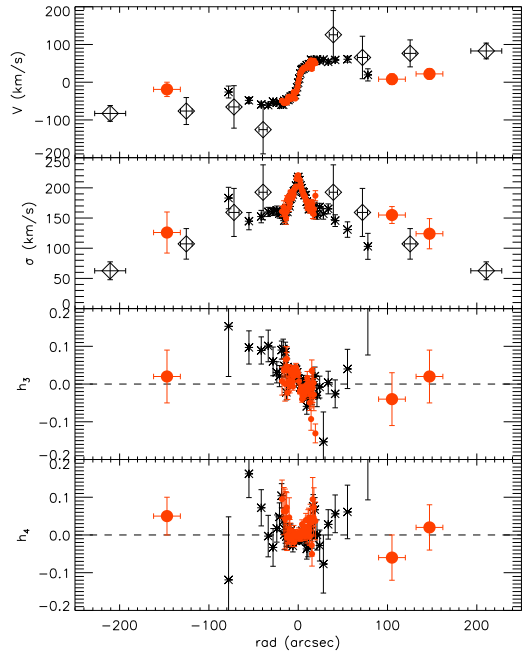,angle=0,width=7.0cm}}
\caption{Kinematic profiles along the major axis of NGC~3379. The
 small red dots denote the central SAURON data from the original
 survey and the big red dots are our data at large radii. For
 comparison, we also plot the long-slit data of Statler \&
 Smecker-Hane (1999) with black stars and the planetary nebulae
 profiles as presented by Coccato et al. (2009) with open
 diamonds. The horizontal error bars for the SAURON data at large
 radii indicate the width of the SAURON field-of-view (30 arcsec),
 while the horizontal error bars of the PNe data indicate the radial
 range containing 68 per cent of the PNe used to calculate the
 kinematics.}
\label{fig:kin_ngc3379}
\end{figure}

\begin{figure}
\centerline{\psfig{figure=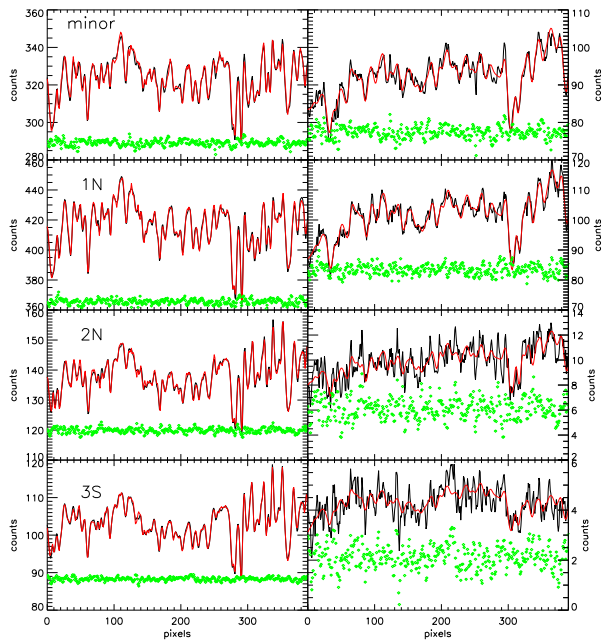,angle=0,width=8.0cm}}
\caption{Same as Figure~\ref{fig:ppxf_3379}, but now for our fields in NGC 821.}
\label{fig:ppxf_821}
\end{figure}

\begin{table*}
\begin{center}
\begin{tabular}{l|c|c|c|c|c|c|c|c|c|c|c|c|c|c}
\hline\hline
Field & $\Delta$x   & $\Delta$y  & $V$ & $\delta V$
 & $\sigma$  & $\delta\sigma$  & $h_3$ & $\delta h_3$ & 
 $h_4$ & $\delta h_4 $ & $V_{\mathrm{mean}}$ & $\delta V_{\mathrm{mean}}$ &
 $\sigma_{\mathrm{mean}}$ & $\delta\sigma_{\mathrm{mean}}$ \\
  & (arcsec) & (arcsec) & (km/s) & (km/s) & (km/s) & (km/s) & & & & & 
  (km/s) & (km/s) & (km/s)  & (km/s) \\

\hline
3S & -77 & -77 &  8 & 14 &  155 & 14 & -0.04 &  0.07 &  -0.06 & 0.06 &
  5 & 11 & 147 & 12 \\
4N & 104 &  104 & -19  & 19 &  126 & 34 &  0.02 & 0.07 &  0.05 & 0.05 & 
  -16 & 19 & 134 & 35 \\
4S & -104 & -104 &  22 & 14 &  124 & 25 &  0.02 & 0.07 &  0.02 & 0.06 & 
 23 & 14 & 127 & 19 \\

\hline

\end{tabular}
\caption{Stellar kinematics at large radii for NGC 3379. $\Delta$x
and $\Delta$y denote the offsets (East and North, respectively) from
the nucleus of the galaxy to the centre of our observed fields. The
quoted errors are 1-$\sigma$ errors, determined from Monte Carlo
simulations.  $V_{\mathrm{mean}}$ and $\sigma_{\mathrm{mean}}$ are
derived by fitting only the first two moments of the LOSVD.}
\label{tab:kin3379}
\end{center}
\end{table*}

In Figure~\ref{fig:kin_ngc3379} we compare our results with the
long-slit data of Statler \& Smecker-Hane
(1999)\nocite{1999AJ....117..839S}. The radial extent of the
absorption line kinematics has been increased by a factor two compared
to the long-slit data, and our data show a smooth continuation of
kinematics out to larger radii. We see little rotation at large radii,
and the dispersion profile remains flat out to 3.5 $R_e$, although there
is a hint that the profile is slightly dropping. We also plot the
kinematic profiles of the planetary nebulae, as presented in Coccato
et al. (2009)\nocite{2009MNRAS.394.1249C}. We folded these profiles to
negative radii, to allow comparison with our data at large radii also
in these ranges. We find a good agreement between the several
datasets, although the rotational velocities found by the planetary
nebulae seem to be a bit higher ($\sim$30 km/s) than our
findings.

\subsection{NGC 821}

The spectra and results for our fields in NGC 821 can be found in
Figure~\ref{fig:ppxf_821} and Table~\ref{tab:kin821}. Unfortunately,
weather conditions did not allow us to obtain enough signal to measure
the LOSVD in all our observed fields, but at each radius one suitable
spectrum could be obtained. The systemic velocity from the
central kinematics (Emsellem et al. 2004\nocite{2004MNRAS.352..721E})
is 1726 $\pm$ 2 km/s (barycentric corrected). This is in agreement
with \vsys\ measured from our two field pairs at 1.9 $R_e$ (2S and
2N), where we find a value of 1704 $\pm$ 25 km/s. Unfortunately, we
could not measure \vsys\ from our field pair at 2.9 $R_e$ due to lack
of signal in field 3N. However, our field at the minor axis of NGC 821
gives an independent determination of \vsys, assuming axisymmetry so
that here the rotational velocity should be zero. We then find \vsys\
= 1710 $\pm$ 7 km/s and \vsys\ = 1716 $\pm$ 7 km/s, depending on
whether we fit up the LOSVD up to $h_4$ or only the first two moments,
respectively. Since all these measurements of \vsys\ agree within the
errors, we fixed \vsys\ such that the velocity measured on the minor
axis is zero.

\begin{table*}
\begin{center}
\begin{tabular}{l|c|c|c|c|c|c|c|c|c|c|c|c|c|c}
\hline\hline
Field & $\Delta$x   & $\Delta$y  & $V$ & $\delta V$
 & $\sigma$  & $\delta\sigma$  & $h_3$ & $\delta h_3$ & 
 $h_4$ & $\delta h_4 $  & $V_{\mathrm{mean}}$ & $\delta V_{\mathrm{mean}}$ &
 $\sigma_{\mathrm{mean}}$ & $\delta\sigma_{\mathrm{mean}}$ \\
  & (arcsec) & (arcsec) & (km/s) & (km/s) & (km/s) & (km/s) & & & & &
 (km/s) & (km/s) & (km/s) & (kms/) \\

\hline
minor & 33 & -21 &  0 & 7 &  171 & 8 &  0.07 & 0.04 & 0.00 & 0.04 &
 0 & 7 & 168 & 7 \\
1N & 19 & 30 &  119 & 10 &  174 & 8 &  0.00 & 0.03 & -0.04 & 0.04 &  
 112 & 9 & 170 & 7  \\
2N & 39 & 63 &  72 & 20 &  131 & 22 & -0.02 & 0.07 & 0.02 & 0.06 &
  64 & 18 & 135 & 19 \\
3S & -60 & -96 &  -1 & 28 &  163 & 43 & -0.02 & 0.06 & 0.04 & 0.06 &
  -8 & 27 & 162 & 38 \\
\hline

\end{tabular}
\caption{Stellar kinematics at large radii for NGC 821, similar to
Table~\ref{tab:kin3379}.}
\label{tab:kin821}
\end{center}
\end{table*}

\begin{figure*}
\begin{tabular}{cc}
\psfig{figure=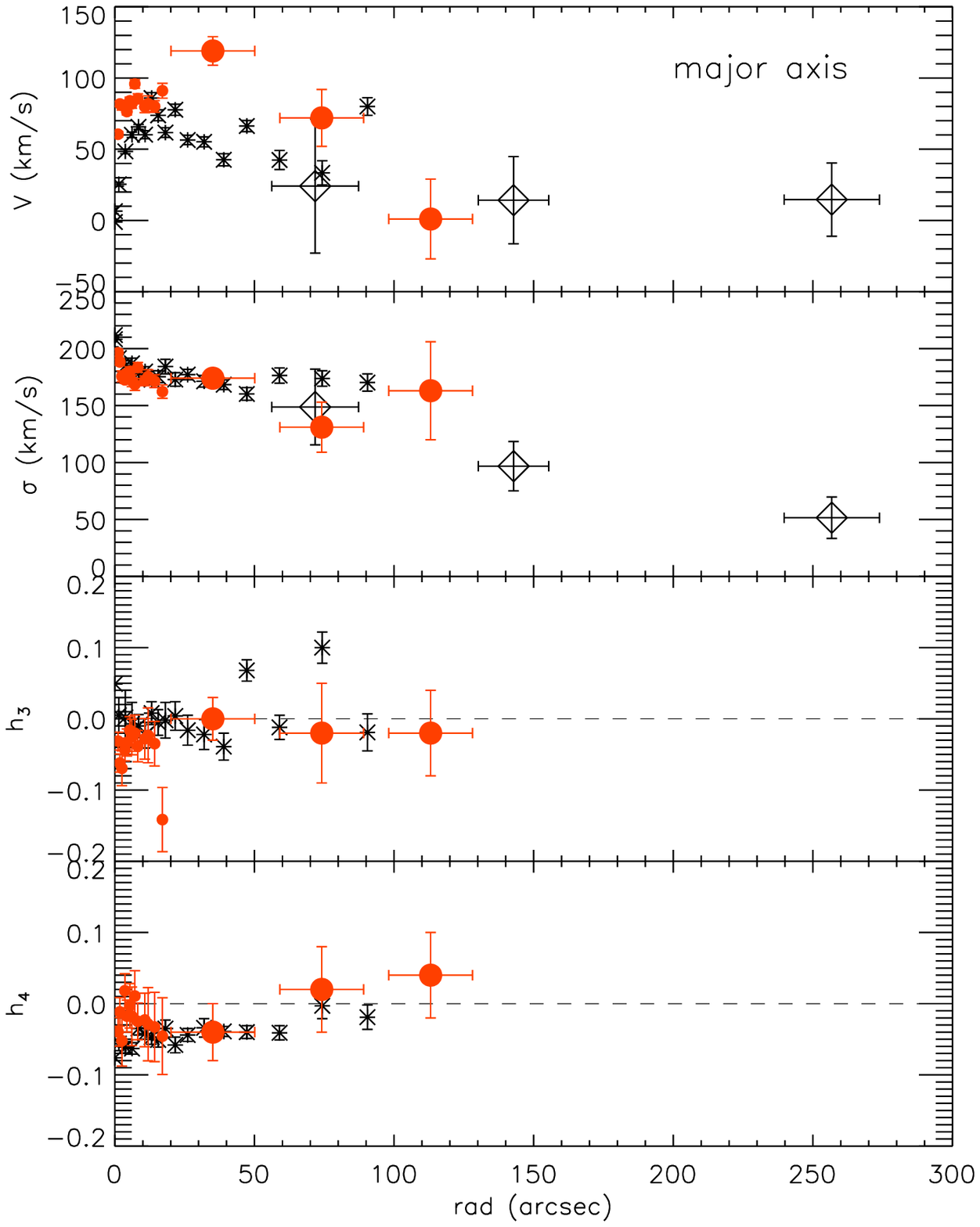,angle=0,width=7.0cm} & \psfig{figure=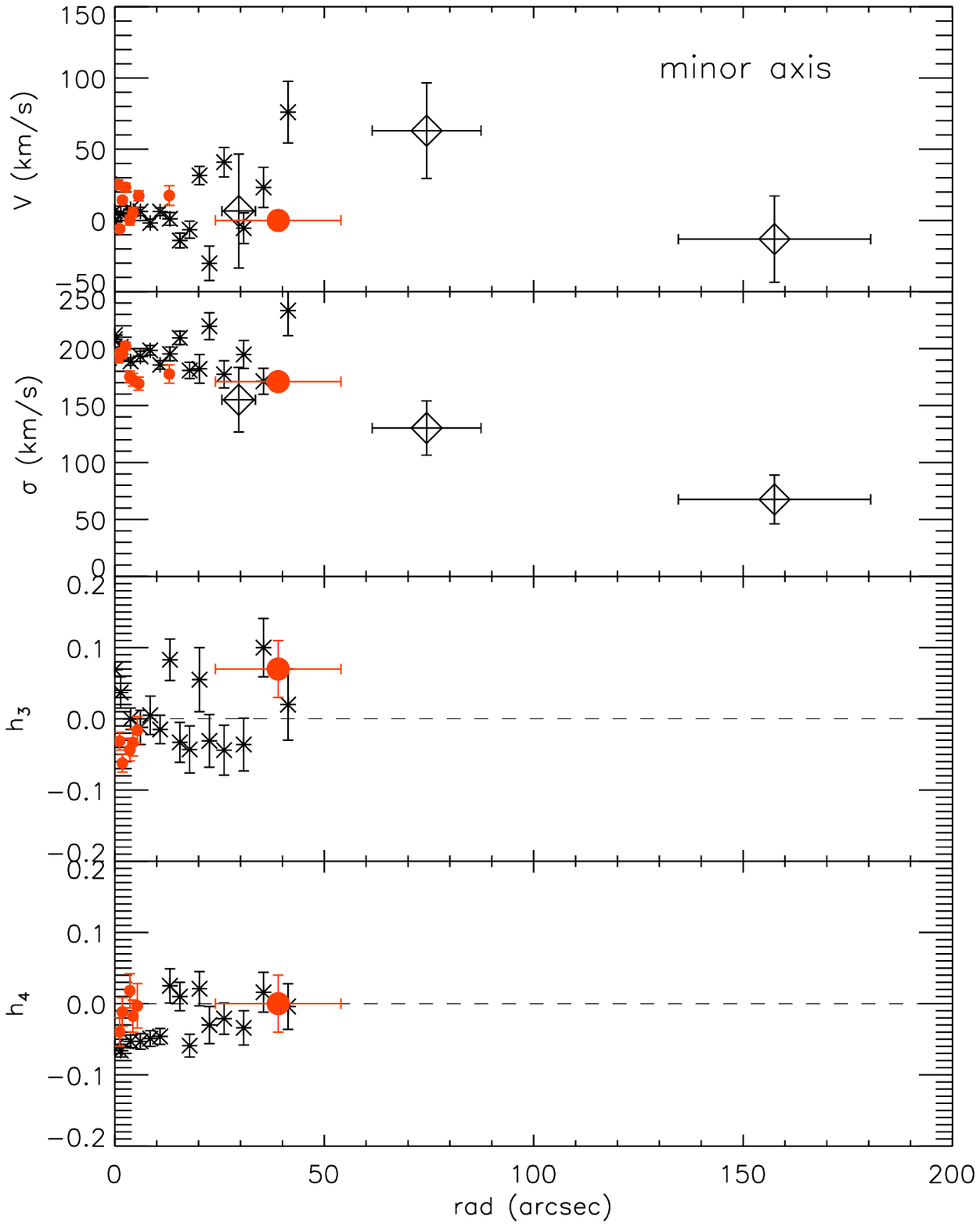,angle=0,width=7.0cm}
\end{tabular}
\caption{Kinematic profiles along the major (left) and minor (right)
axis of NGC~821. As in Figure~\ref{fig:kin_ngc3379}, small red dots
are the central SAURON data of the original survey, big red dots are
our data at large radii and open diamonds denote planetary nebulae
kinematics from Coccato et al. (2009). The black stars are long-slit
stellar kinematics from Forestell \& Gebhardt (2008). Horizontal error
bars are as described in Figure~\ref{fig:kin_ngc3379}.}
\label{fig:kin_ngc821}
\end{figure*}

Figure~\ref{fig:kin_ngc821} compares our result with the long-slit
data of Forestell \& Gebhardt (2008)\nocite{2008arXiv0803.3626F}. We
note that especially in the centre there seem to be some deviations
between this dataset and the SAURON data. Differences in sampling
between the two datasets (nearly-round bins for the SAURON data versus
long, elongated bins for the long slit data) could in principle result
in differently-weighted values, but we expect that this effect is
small. The most likely cause for the observed deviations are
differences in fitted wavelength interval and stellar templates.  Also
shown are the results from Coccato et
al. (2009)\nocite{2009MNRAS.394.1249C}, who re-analysed the planetary
nebulae data that were previously presented by Romanowsky et
al. (2003)\nocite{2003Sci...301.1696R}. These data are in good
agreement with our stellar kinematics at large radii.

%%%%%%%%%%%%%%%%%%%%%%%%%%%%%%%%%%%%%%%%%%%%%%%%%%%%%%%%%%%%%%%%%%

\section{Line strengths}

To measure the absorption line strengths in our spectra, we follow the
procedure outlined in Kuntschner et
al. (2006)\nocite{2006MNRAS.369..497K}.
%\subsection{Method}
Briefly, we can measure three Lick indices in the SAURON wavelength
range: \Hb, Fe5015 and \Mgb, defined in Table~2 of Kuntschner et
al. (2006)\nocite{2006MNRAS.369..497K}. We do not need to remove
emission lines from our spectra, since our galaxies show no sign of
ionised gas outside the central regions (Sarzi et
al. 2006\nocite{2006MNRAS.366.1151S}). To calibrate to the Lick/IDS
system, we broaden our spectra to an instrumental resolution of
$\sigma$ = 212 km s$^{-1}$. We correct our measured line strengths for
the broadening caused by the LOSVD and we apply the offsets from
Table~3 in Kuntschner et al. (2006)\nocite{2006MNRAS.369..497K} to
correct for differences in the shape of the continuum, since the
Lick/IDS spectra have not been flux calibrated.

We noticed that the continuum shape blue-ward of the \Hb\ feature in
our spectra was affected by bad flatfielding or wrong sky
subtraction. We therefore applied the same continuum correction as was
done in Kuntschner et al. (2006)\nocite{2006MNRAS.366.1151S} and
fitted an 11$^{\mathrm{th}}$ order multiplicative polynomial together
with an optimal stellar template to our spectra. The template had been
broadened with the observed LOSVD. We then divided our observed
spectrum by this polynomial, before determining the \Hb\ index.

\subsection{Results}

We present our results for NGC 3379 and NGC 821 in
Table~\ref{tab:lick}. Errors are estimated with Monte Carlo
simulations, where we add noise to our spectra (defined by the
difference between our observed and fitted spectra, see
Figures~\ref{fig:ppxf_3379} and \ref{fig:ppxf_821}) and vary the
bandpasses of the index and the continuum to take the uncertainty in
radial velocity into account. 

A possible source of contamination for our line strength measurements
is the \Oiii\ emission of planetary nebulae. From the catalog of
Douglas et al. (2007)\nocite{2007ApJ...664..257D}, each of our fields
in NGC~3379 contains one to two planetary nebulae. Following Sarzi et
al. (2005)\nocite{2006MNRAS.366.1151S}, their Equation 1, we should be
sensitive to emission lines with EW $>$ 0.7 \AA. We carefully checked
the individual spectra in each cube, prior to co-adding, and found no
evidence for such emission.

\begin{table}
\begin{center}
\begin{tabular}{l|c|c|c|c|c|c|}
\hline\hline
Field & \Hb & error & Fe5015 & error & \Mgb & error \\
 & (\AA) & (\AA)  & (\AA) & (\AA)  & (\AA) & (\AA) \\
\hline
NGC 821 & & & & & & \\
minor & 1.61  & 0.22 & 3.80 & 0.28  & 3.13 & 0.16 \\
1N    & 1.30  & 0.20 & 4.11 & 0.24  & 3.45 & 0.17 \\
2N    & 1.19  & 0.81 & 5.16 & 1.37  & 5.47 & 0.64 \\
3S    & 0.00 & 1.19  & 12.0 & 1.71  &6.43 & 0.76 \\
\hline
NGC 3379 & & & & & & \\
3S & 1.49 & 0.45 & 2.45 & 0.58 & 3.42 & 0.26 \\
4N & 2.13 & 0.92 & 2.66 & 0.86 & 3.05 & 0.45 \\
4S & 1.17 & 0.51 & 2.46 & 0.66 & 3.61 & 0.43 \\
\hline

\end{tabular}
\caption{Line strengths indices in our observed fields in NGC
821 and NGC 3379.}
\label{tab:lick}
\end{center}
\end{table}

\begin{figure*}
\begin{tabular}{cc}
\psfig{figure=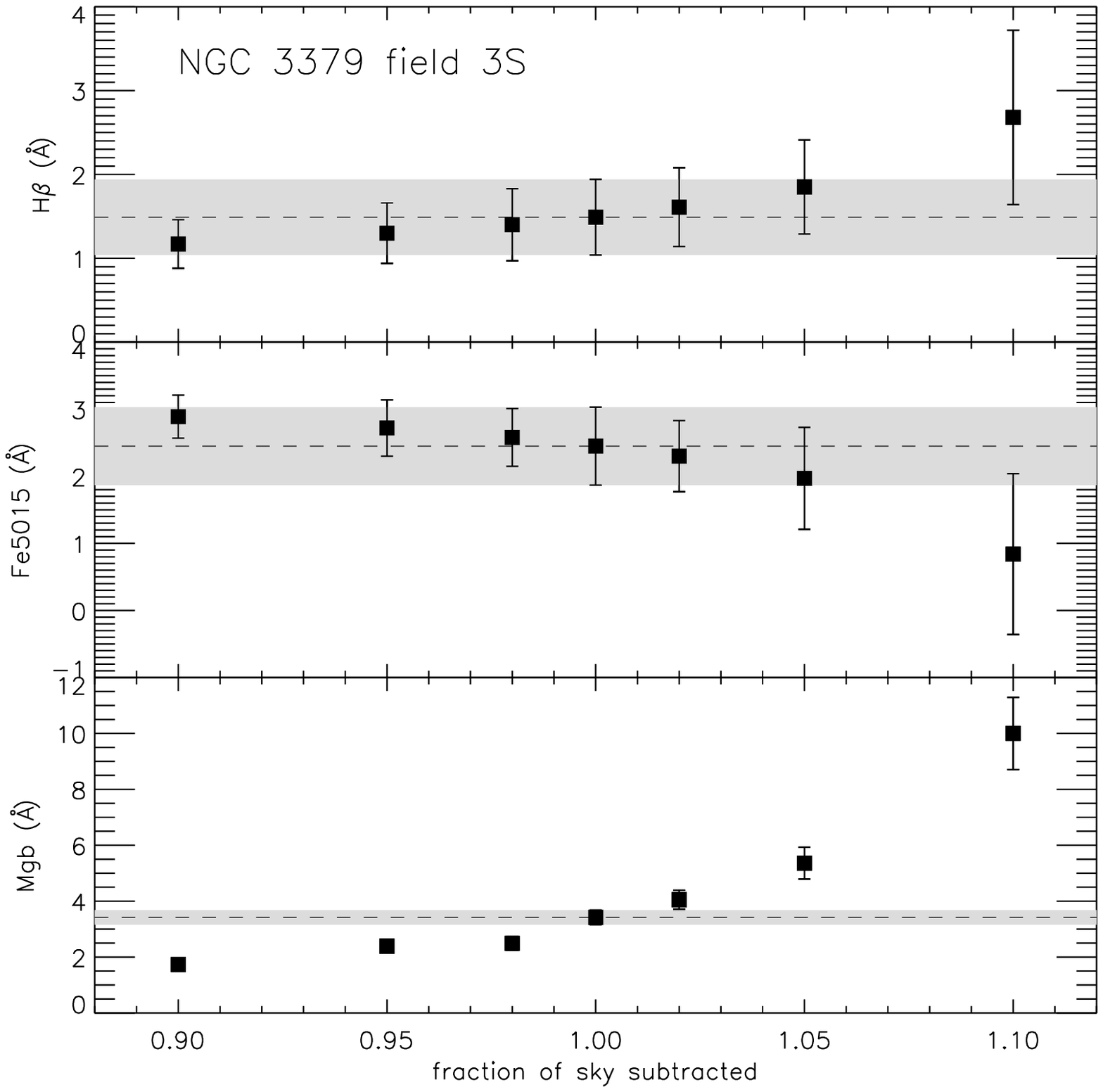,angle=0,width=8.0cm} & \psfig{figure=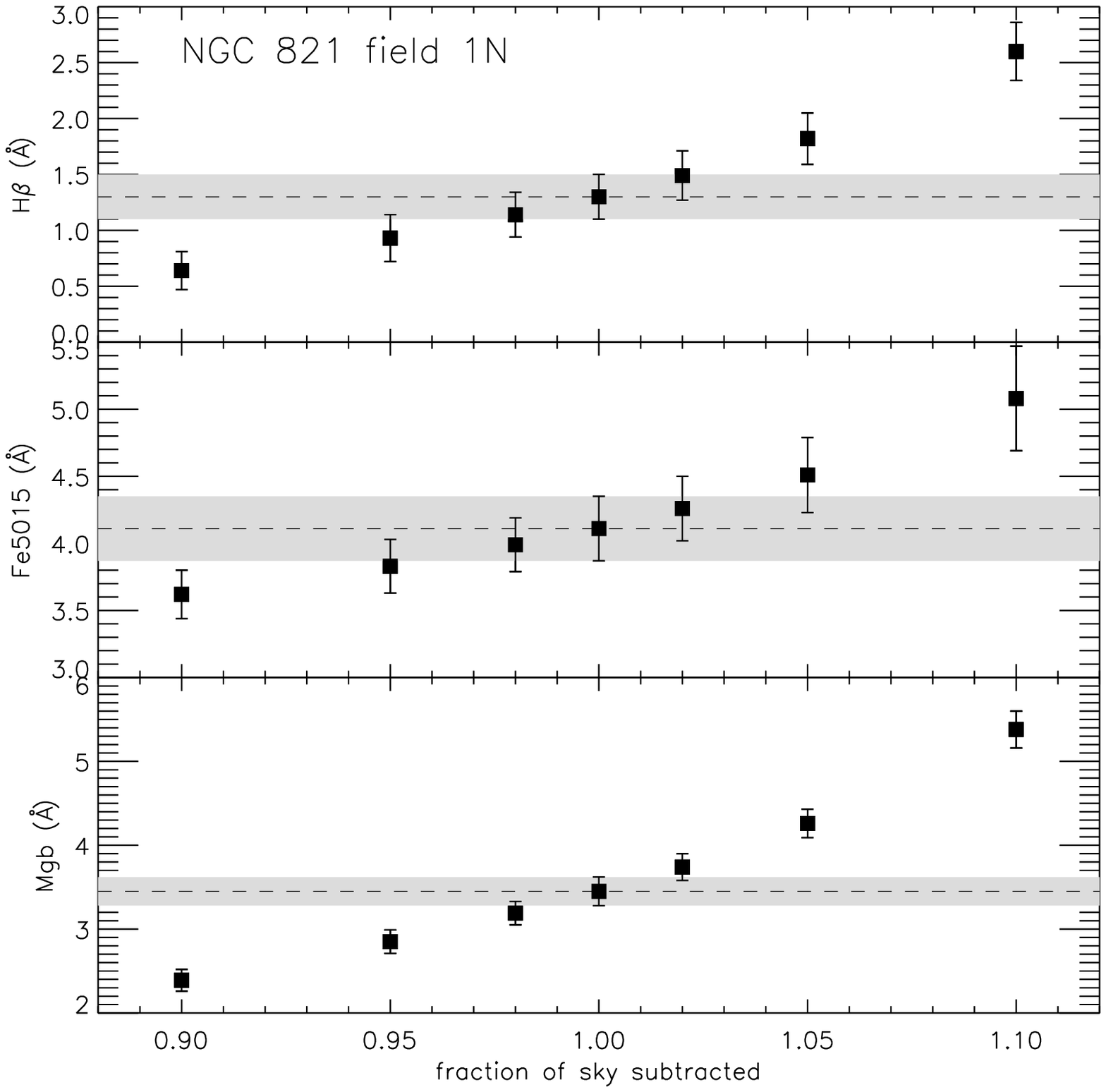,angle=0,width=8.0cm}
\end{tabular}

\caption{The effect of sky subtraction on our line strength measurements, for the field 3S in NGC~3379 (left) and the 1N field in NGC~821 (right). The $x$-axis denotes the fraction of sky subtracted. The dashed line indicates the line index when the sky is totally removed, while the grey box denotes the 1-$\sigma$ interval. \Hb\ and Fe5015 are robust against an error in sky subtraction up to 5 per cent, though oversubtracting is more severe than undersubtracting. \Mgb\ is very sensitive to sky subtraction errors.}

\label{fig:lick_sky}
\end{figure*}

We also investigated the effect of sky subtraction on the
determinations of line strength. By varying the amount of sky that we
subtracted from our spectra, we found that our measurements of \Hb\ and
Fe5015 are quite robust against an error in sky subtraction, even if
the skylevels are varied by 10 per cent. \Mgb\ is however already
severely affected by a sky subtraction error of only two per cent (see
Figure~\ref{fig:lick_sky}). This may be due to the proximity of the
solar \Mgb\ absorption feature present in the sky spectrum. Our Fe5015
and \Mgb\ indices at 2 and 3 $R_e$ in NGC 821 are unrealistically
high, and therefore could be suffering from this problem, although
variations in the continuum shape of the spectrum also play a role
(see next section). The indices at 1 $R_e$ in this galaxy are less
affected, and are in agreement with Proctor et
al. (2005)\nocite{2005MNRAS.362..857P}. They determined line strengths
from long-slit data in this galaxy out to 1 $R_e$.  We are not aware
of studies in the literature where line strengths have been determined
outside 1 $R_e$ for these galaxies, to compare our results with.

\subsection{Line strength gradients}

\begin{figure*}
\begin{tabular}{cc}
\psfig{figure=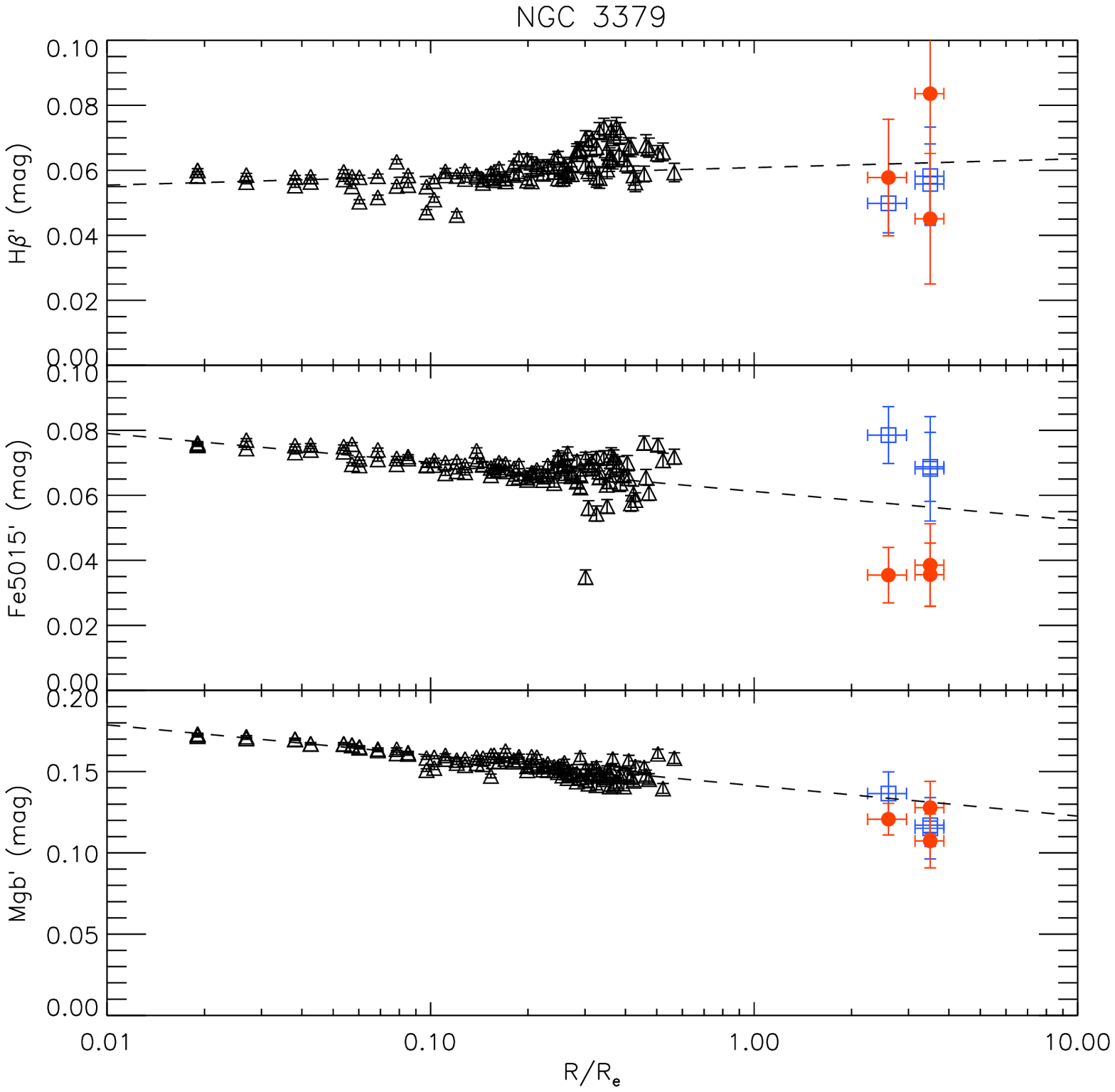,angle=0,width=8.0cm} & \psfig{figure=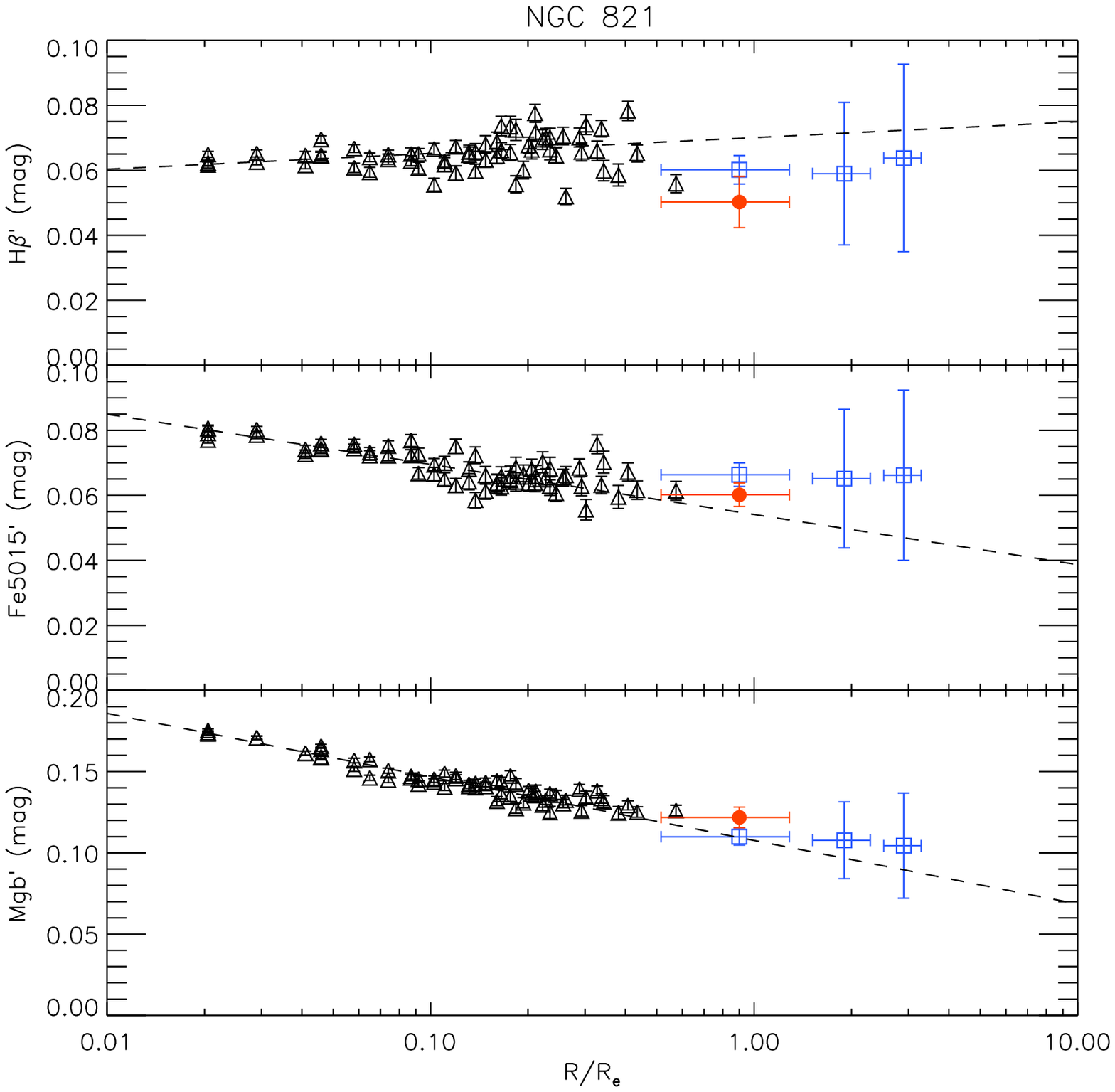,angle=0,width=8.2cm}
\end{tabular}
\caption{Line strength indices \Hb, Fe5015 and \Mgb\ (in magnitudes) as a function of radius. Left panel: NGC 3379, right panel: NGC 821. Black triangles show indices measured on the SAURON central field, taken from a slit positioned on the major axis of the galaxy, with a width of 2 arcsec. Red filled dots show our large radii results, and the horizontal error bars denote the size of the SAURON field of view. The points at 2 and 3 $R_e$ in NGC 821 cannot be considered reliable, probably because of continuum shape problems, and therefore are not shown here. The blue open squares denote our measurements directly from templates of single stellar population models, see text for details. Overplotted is the best fitting linear relation between index and log(radius), determined within $R_e/4$.}
\label{fig:lick_rad}
\end{figure*}

Line strength gradients are key instruments to study the formation
history of galaxies, since different formation models predict
different gradients. Monolithic collapse models (Carlberg
1984\nocite{1984ApJ...286..403C}) produce steep metallicity gradients
and a metal-rich nucleus, while hierarchical models show shallower
gradients, as mergers dilute existing line strength gradients (White
1980\nocite{1980MNRAS.191P...1W}). However, subsequent starformation
induced by gas-rich mergers can steepen the gradients again, at least
in the central regions (Hopkins et
al. 2009a\nocite{2009ApJS..181..135H}).

Our measurements allow us to study line strength gradients out to
large radii. Many previous studies of galaxy samples (e.g. Davies,
Sadler \& Peletier 1993\nocite{1993MNRAS.262..650D}; Kuntschner et
al. 2006\nocite{2006MNRAS.366.1151S}; S\'anchez-Bl\'azquez et
al. 2007\nocite{2007MNRAS.377..759S}) have been limited to $\sim$ 1
$R_e$, but showed that typically for galaxies without contributions
from young stellar populations, \Hb\ remains flat or increases
slightly with radius, while the metal line strengths have negative
gradients.

To calculate the line strength gradients in our two galaxies, we
follow Kuntschner et al. (2006)\nocite{2006MNRAS.366.1151S} and first
convert our indices from \AA\ to magnitudes:

\begin{equation}
\mathrm{index'} = -2.5\log\large( 1 - \frac{\mathrm{index}}{\Delta\lambda}\large),
\end{equation}

\noindent
with $\Delta\lambda$ the width of the index bandpass in \AA. Indices
expressed in magnitudes will from now on be indicated by a prime sign
[$^{\prime}$]. Line strength gradients are then defined as:

\begin{equation}
\Delta \mathrm{index'} = \frac{\delta \mathrm{index'}}{\delta \log (R/R_e)}.
\end{equation}

\noindent
We show in Figure~\ref{fig:lick_rad} our line strength indices in
magnitudes as a function of radius, with red dots. Also plotted are
the values for the central SAURON fields (Kuntschner et
al. 2006\nocite{2006MNRAS.366.1151S}), which fall within a 2 arcsec
wide slit positioned along the major axis of the galaxy. We fitted a
linear relation between each index and $\log$(radius) within $R_e/4$,
to determine the gradient. These gradients are overplotted in the
figure, and are compared to our measurements at large radii.

With our new data, we confirm in NGC 3379 the trends mentioned above,
that \Hb\ mildly increases and Fe5015 and \Mgb\ decrease with
radius. This indicates that the metallicity in the galaxy keeps
decreasing linearly with log(radius) when going from the centre out to
larger radii. The gradients for \Hb\ and \Mgb\ seem to remain constant
out to 4 $R_e$, but Fe5015 falls off more steeply. This would imply an
unrealistically high $\alpha$-enhancement at large radii. To check
whether our Fe5015 measurements are affected by continuum shape
problems, we did the following test. We remeasured the kinematics in
our spectra with pPXF, but this time used the stellar model library of
Vazdekis et al. (2007)\nocite{2007IAUS..241..133V} based on the MILES
library (-1.68 $<$ $Z/H$ $<$ 0.0; 1 Gyr $<$ age $<$ 17.78 Gyr) and
multiplicative instead of additive polynomials, to allow for a varying
continuum shape. The resulting kinematics were consistent with our
previous finding. The line strengths were then measured on the
composite best-fitting stellar template, without the multiplicative
polynomials. The results are also shown in Figure~\ref{fig:lick_rad},
as blue squares. For \Hb\ and \Mgb\ both the direct measurements and
the measurements on the stellar templates give similar results, but
the values for Fe5015 are higher and consistent with the observed
gradient in the central part of the galaxy. We therefore conclude that
indeed our measurements for Fe5015 suffer from changes in continuum
shape, but that our values for \Hb\ and \Mgb\ are robust. For NGC 821
we see a similar effect.

Our observation that the line strength gradients are constant out to
large radii provides constraints for the merger and star formation
history of the galaxy. After a gas-rich merger, star formation in the
central regions of the remnant is expected to steepen the gradients, while at
larger radii violent relaxation flattens them, though over time ($\sim
3$ Gyr) the steep gradients in the central part get weaker (Hopkins et
al. 2009a\nocite{2009ApJS..181..135H}; Hopkins et
al. 2009b\nocite{2009ApJS..181..486H}).

\subsection{Stellar population models}

Using the stellar population models of Thomas, Maraston \& Bender
(2003)\nocite{2003MNRAS.339..897T} and Schiavon
(2007)\nocite{2007ApJS..171..146S}, we explore the single stellar
population (SSP) equivalent age and metallicity in NGC 3379 and NGC
821 at large radii.  Following the aproach of Thomas et
al. (2003)\nocite{2003MNRAS.339..897T} we define an abundance ratio
insensitive metallicity index similar to their [MgFe]' index by using
the \Mgb\ and Fe5015 indices available from SAURON. [MgFe50]' is
defined as:

\begin{equation}
\mathrm{[MgFe50]'} = \frac{0.69\times \mathrm{Mg}b + \mathrm{Fe5015}}{2}.
\end{equation}

\begin{figure*}
\begin{tabular}{c}
\psfig{figure=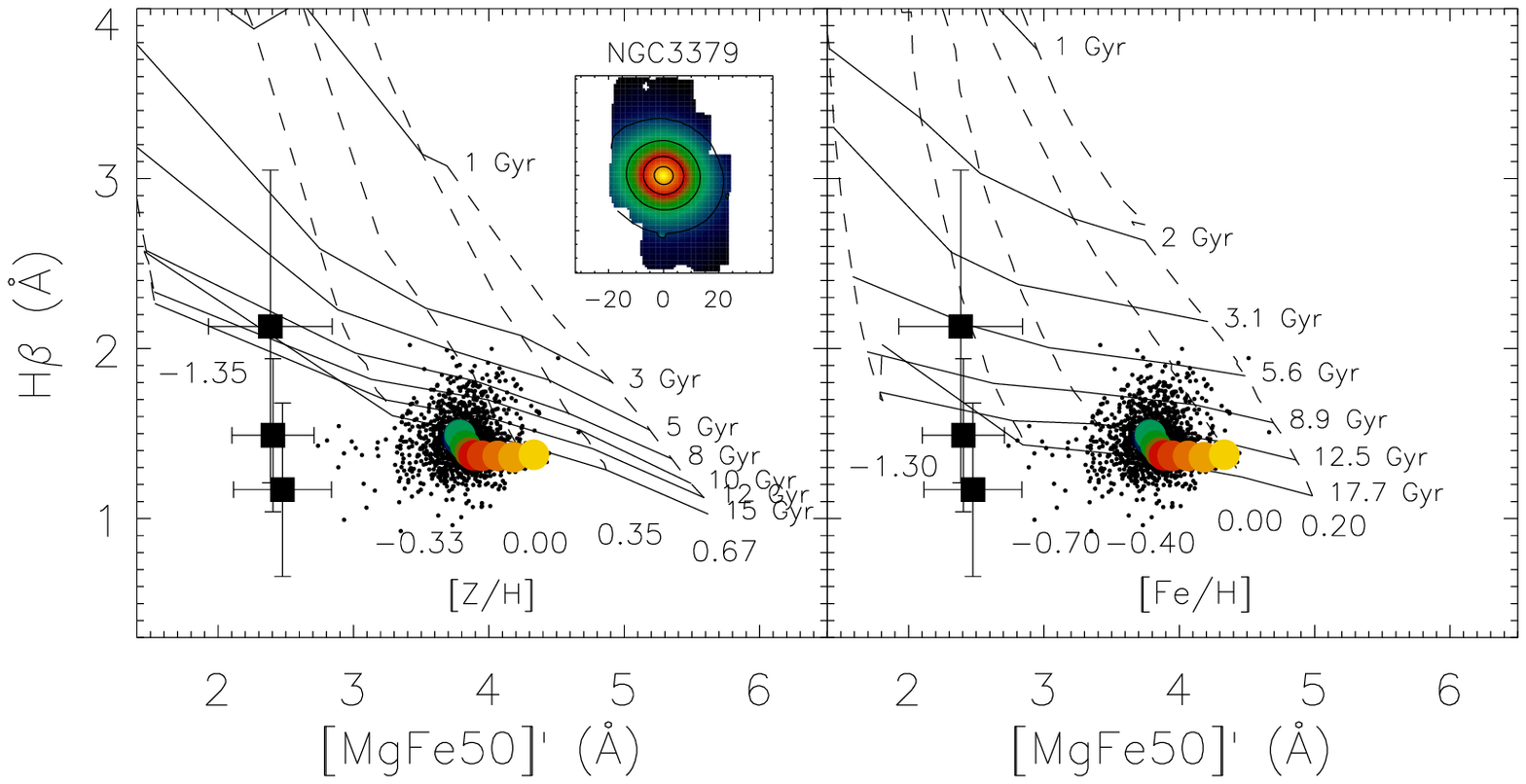,angle=0,width=14.0cm} \\
\psfig{figure=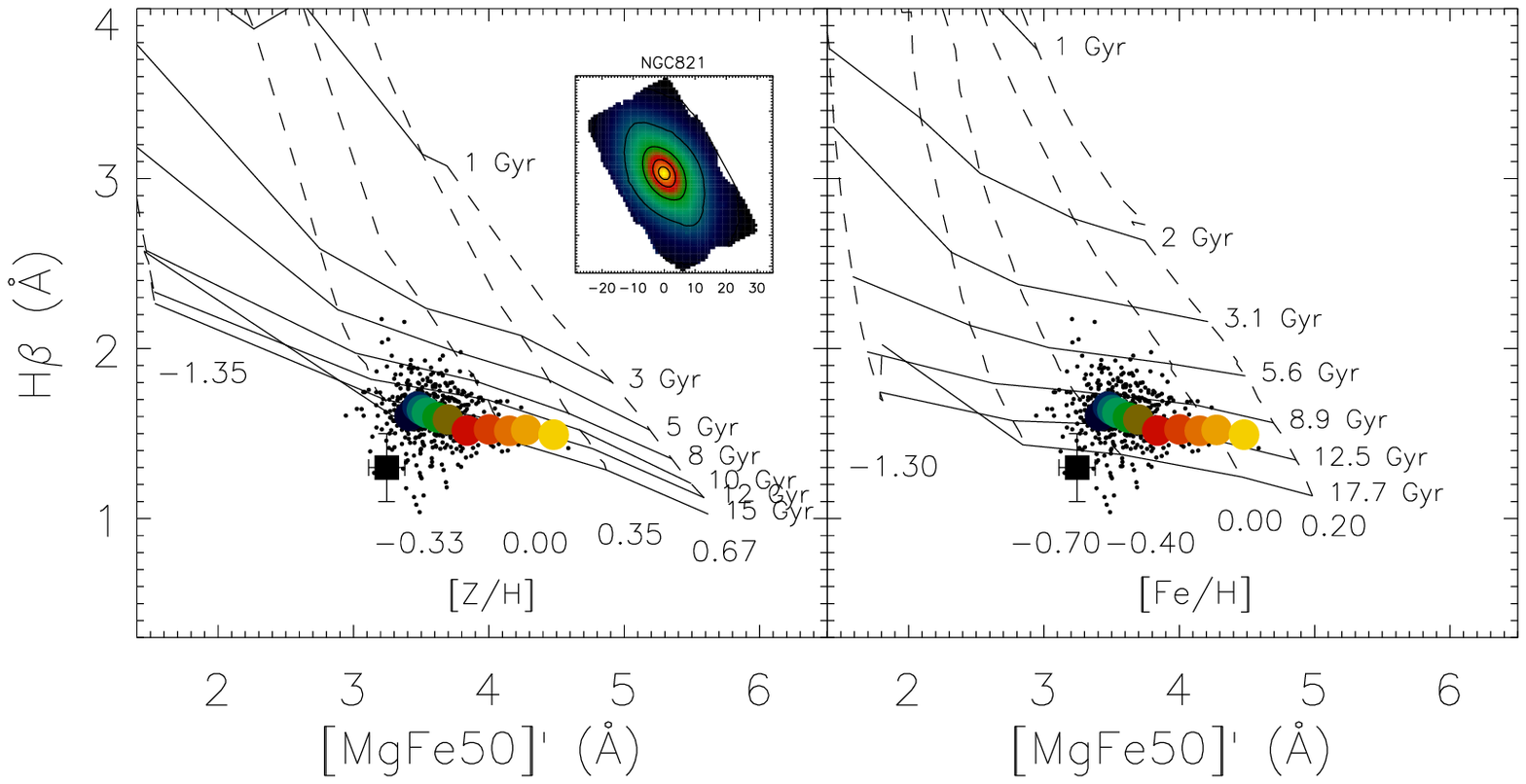,angle=0,width=14.0cm}
\end{tabular}

\caption{\Hb\ index against [MgFe50]$^{\prime}$ (both in \AA) in NGC
3379 (top) and NGC 821 (bottom). Left panels show the Thomas et
al. (2003) stellar population models (solid and dashed lines) and
right panels the Schiavon (2007) models. Black dots indicate
measurements from the SAURON central field, while the coloured dots
are averaged along isophotes (see inset for colour coding). The black
filled squares show the data at large radii, revealing an old (12 Gyr)
and metal-poor (below 20 per cent solar metallicity) population for
both models, though the uncertainties in these values are large.}

\label{fig:lick_model}
\end{figure*}

\noindent
The scaling factor for the \Mgb\ index was optimized such that the
mean difference between solar and non-solar ratio model predictions
from Thomas et al. (2003)\nocite{2003MNRAS.339..897T} is zero. In
Figure~\ref{fig:lick_model} we plot this index versus \Hb\, and
compare with the models. Even though the models of Thomas et
al. (2003)\nocite{2003MNRAS.339..897T} and Schiavon
(2007)\nocite{2007ApJS..171..146S} are constructed independently,
using different stellar libraries and fitting functions, they give the
same results for our data at large radii. For NGC 3379 we find that at
3 - 4 $R_e$, the stellar population is consistent with an old (12 Gyr)
population, and metal-poor, with [Z/H] slightly below 20 per cent of
the solar metallicity. We note however that the uncertainty in these
values is rather large. For NGC 821, we find that at 1 $R_e$ the
stellar population is of the same age and metallicity range as
obtained from the outer bins of the SAURON central field ($\sim$ 0.6
$R_e$).

Stellar population models predict a decrease in stellar mass-to-light
ratio $M_*/L$ if the metallicity of the stellar population decreases,
since the stars then become bluer and therefore brighter in the
optical. For instance, for a change in metallicity from 0.0 to -0.84
at a constant age of 10 Gyr (consistent with our observervations),
$M_*/L$ decreases by about 23 per cent in the models of Maraston
(2005)\nocite{2005MNRAS.362..799M}. However, for increasing stellar
age at constant metallicity, $M_*/L$ also increases. Given this
degeneracy and uncertainties, we adopt a constant stellar $M/L$ while
constructing the dynamical mass models that we present in the next
section. Furthermore, $M_*/L$ depends strongly on \Hb\ (see e.g.
Cappellari et al. 2006\nocite{2006MNRAS.366.1126C}). Since the \Hb\
profiles in our galaxies are nearly flat out to large radii,
variations in $M_*/L$ are most likely small.

%%%%%%%%%%%%%%%%%%%%%%%%%%%%%%%%%%%%%%%%%%%%%%%%%%%%%%%%%%%%%%%%%%

\section{Dynamical models}

To explore whether our data of NGC~3379 and NGC~821 are consistent
with a dark matter halo, we model these galaxies with the triaxial
Schwarzschild code presented by van den Bosch et
al. (2008).\nocite{2008MNRAS.385..647V} Orbits are calculated within
an a priori specified triaxial potential, and a superposition of
orbits is determined that best fits the observed kinematics, using the
observed surface brightness and self-consistency as constraints. This
way, the viewing directions, stellar mass-to-light ratio $M_*/L$,
central black hole mass $M_\bullet$ as well as the internal orbital
structure can be determined (see also van de Ven, de Zeeuw \& van den
Bosch 2008\nocite{2008MNRAS.385..614V}; van den Bosch \& van de Ven
2008\nocite{2008arXiv0811.3474V}). Within the limits of the observed
photometry and LOSVD, the velocity anisotropy of the galaxy can vary
freely.

\subsection{Including a dark halo}

We extended the triaxial Schwarzschild code by adding a dark matter
halo to the gravitational potential, which already consisted of a
stellar contribution inferred from the observed photometry and a
Plummer potential for the contribution of the central black hole (see
van den Bosch et al. 2008\nocite{2008MNRAS.385..647V} for more
details). We parametrise the dark halo with a spherical NFW profile
(Navarro, Frenk \& White 1996\nocite{1996ApJ...462..563N}):

\begin{equation}
\label{eq:NWFprof}
\rho(r) = \frac{\rho_s}{r/r_s\large(1 + r/r_s \large)^2},  
\end{equation}

\noindent
with $\rho_s$ the characteristic density of the halo and $r_s$ a
characteristic radius. The potential generated by this density
distribution is then given by:

\begin{equation}
\label{eq:NFWpot}
\Phi(r) = -4\pi G \rho_s r_s^2 \frac{r_s}{r} \ln \large( 1 + \frac{r}{r_s} \large),
\end{equation}

\noindent
where $G$ is the gravitational constant. 

We are interested in the presence and acceptable mass range of the
halo; modeling the shape of the halo in detail is beyond the scope of
this paper. We therefore fix the concentration $c = r_{200}/r_s$ of
the halo to 10, as indicated by cosmological simulations (Bullock et
al. 2001\nocite{2001MNRAS.321..559B}). Here $r_{200}$ is the radius of
the halo within which the mean density has dropped to 200 times the
critical density $\rho_{\mathrm{crit}}$. It can be shown that

\begin{equation}
\label{eq:c}
\frac{\rho_s}{\rho_{\mathrm{crit}}} = \frac{200}{3} \frac{c^3}
  {\ln(1 + c) - c/(1+c)},
\end{equation}

\noindent
such that $c$ determines $\rho_s$ in the potential of
Equation~\ref{eq:NFWpot}. We vary the halo mass, 

\begin{equation}
\label{eq:M200}
M_{200} = 4\pi\rho_s r_s^3 \big[ \ln(1+c) - \frac{c}{1+c} \big],
\end{equation}

\noindent
which is the enclosed mass within $r_{200}$, yielding $r_s$. We
only fit for $M_{200}$ and keep $M_*/L$ fixed to the maximum value
allowed by the data, therefore fitting a minimal halo or maximal
spheroid\footnote{Comparable to the maximal disc model that is often
invoked when modeling dark haloes of spiral galaxies}. The influence
of the black hole is negligible at large radii and therefore we also
do not fit for $M_\bullet$ but keep its value fixed.

\subsection{Modelling NGC~3379 and NGC~821}

NGC~3379 has a small kinematic misalignment of $5 \pm 3^\circ$
(e.g. Statler \& Smecker-Hane 1999\nocite{1999AJ....117..839S}) and
shows signs of a kinematic twist (Krajnovi\'c et
al. 2008)\nocite{2008MNRAS.390...93K}, which indicate that the galaxy
is not perfectly axisymmetric (see also Capaccioli et
al. 1991\nocite{1991ApJ...371..535C}; Statler
2001\nocite{2001AJ....121..244S}). Van den Bosch \& de Zeeuw
(2009)\nocite{2009MNRAS.subm} therefore constructed triaxial
Schwarzschild models of this galaxy. Their best-fitting triaxial model
(axis ratios $p$ = 0.95, $q$ = 0.81) is nearly face-on, with an
inclination of $\sim40^\circ$. This model is nearly spherical in the
centre and becomes close to oblate at large radii, which is in
agreement with Statler (2001)\nocite{2001AJ....121..244S} and De
Lorenzi et al. (2009)\nocite{2009MNRAS.395...76D}.

\begin{figure}
\centerline{\psfig{figure=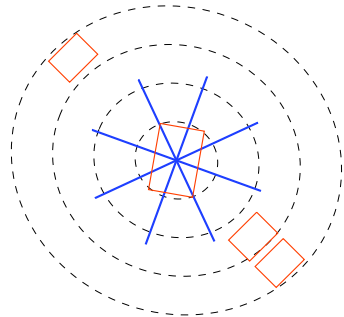,angle=0,width=5.0cm}}
\caption{Overview of the datasets included in our dynamical model of
NGC~3379. The dotted lines denote isophotes of the galaxy at 1, 2, 3
and 4 $R_e$. The (red) box in the centre denotes the central SAURON
field (Emsellem et al. 2004), while the other boxes show the positions
of our data at large radii. The (blue) solid lines show the
orientation and extent of the long-slit data (Statler \& Smecker-Hane
1999).}
\label{fig:foot}
\end{figure}

We adopt the best-fit model of van den Bosch \& de Zeeuw
(2009)\nocite{2009MNRAS.subm} for our halo modeling. The model has a
black hole mass of $M_\bullet = 4.0 \times 10^8 M_\odot$ and a
dynamical mass-to-light ratio $M/L_I$ = 2.9 $M_\odot/L_{I,\odot}$,
which we adopt as the maximally allowed stellar mass-to-light ratio
$M_*/L$. The stellar potential is represented by a Multi-Gaussian
Expansion (MGE, Emsellem, Monnet \& Bacon
1994\nocite{1994A&A...285..723E}; Cappellari
2002)\nocite{2002MNRAS.333..400C} model based upon $I$-band
imaging\footnote{Obtained with the 1.3-m McGraw-Gill Telescope of the
  MDM Observatory, Kitt Peak (Falc\'on-Barroso et al. in prep)},
presented by Cappellari et al. (2006)\nocite{2006MNRAS.366.1126C}. We
use the same central SAURON kinematics (up to $h_6$) extracted with
the MILES library (see Shapiro et
al. 2006\nocite{2006MNRAS.370..559S}; Cappellari et
al. 2007\nocite{2007MNRAS.379..418C}; van den Bosch \& de Zeeuw
2009\nocite{2009MNRAS.subm}) to cover the inner $R_e$ of NGC~3379 and
add our measure points at large radii. For extra spatial coverage, we
also include the symmetrized data of the four long-slit observations
of Statler \& Smecker-Hane (1999)\nocite{1999AJ....117..839S} (see
Figure~\ref{fig:foot}). The large mosaic of central SAURON data
provides many more kinematic constraints in the central region than
the four slits. We therefore exclude the inner 20 arcsec of each slit
from the fit. The planetary nebulae measurements are not based on
integrated stellar light, as our SAURON and long-slit data, and
therefore for consistency not included.

We use a similar procedure to construct mass models of NGC~821. This
galaxy is a very flattened system (E6) and shows no signs of
non-axisymmetry in its central part, even though planetary nebulae
kinematics suggest twists at larger radii (Coccato et
al. 2009\nocite{2009MNRAS.394.1249C}). Our dataset however provides no
constraints on the behaviour of the velocity field at these scales. We
therefore deproject this galaxy axisymmetrically, assuming an edge-on
view and use the triaxial code in the axisymmetric limit. We adopt the
edge-on MGE model of NGC~821 based upon ground-based $R$-band
photometry\footnote{Obtained with the 0.9-m telescope at the Cerro
Tololo Inter-American Observatory (CTIO), Chile} and space-based
imaging with HST/WFPC2 in F814W-band, as in McDermid
(2002)\nocite{2002PhD} and Cappellari et
al. (2006)\nocite{2006MNRAS.366.1126C}. We first construct a model
without a dark halo, based only on kinematics (up to $h_6$) of the
central SAURON dataset, extracted with the MILES library (Cappellari
et al. 2007\nocite{2007MNRAS.379..418C}). Our best fit has a stellar
mass-to-light ratio of 2.5 $M_\odot/L_{I,\odot}$ with a black hole
mass of $M_\bullet = 2.1 \times 10^8 M_\odot$ from the
$M_\bullet$-$\sigma$ relation. We then add a dark NFW halo to this
model, with a fixed concentration of $c=10$. The SAURON central
dataset is used to cover the central part of the galaxy, while our
four measurepoints at large radii are included to cover the
outerparts. To avoid a possible conflict between the central SAURON
data and the long-slit data of Forestell \& Gebhardt
(2008)\nocite{2008arXiv0803.3626F}, we decided to not include the
latter in our model, as there seem to be some deviations between the
two datasets (see Figure~\ref{fig:kin_ngc821}).

\begin{figure}
\psfig{figure=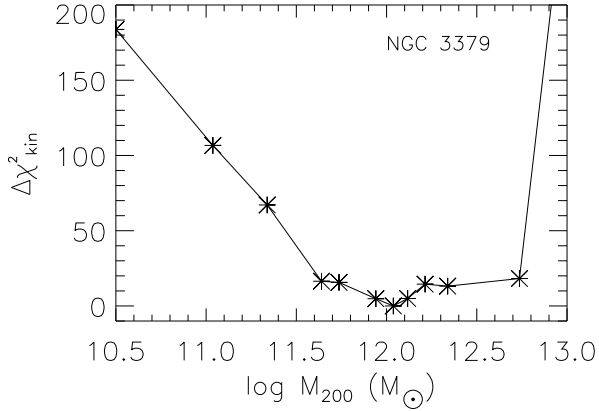,angle=0,width=8.0cm}
\caption{$\Delta \chi^2$ levels for the Schwarzschild models we ran
for NGC~3379. The only free parameter in these models was $M_{200}$
(total halo mass). The model without dark halo is relocated at $3
\times 10^{10} M_\odot$, to place it within the plot.}
\label{fig:chi2_3379}
\end{figure}

\begin{figure}
\psfig{figure=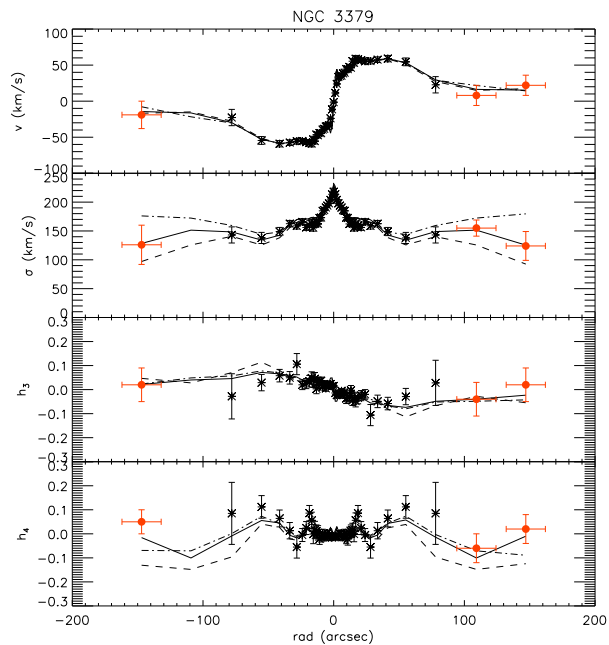,angle=0,width=8.0cm}
\caption{Best-fit model for NGC~3379 (bold line, $M_{200} = 1.0 \times
10^{12} M_\odot$) compared to our datapoints at large radii (red {\bf
filled circles}) and long-slit data on the major axis (black
stars). Also indicated are a model without halo (dashed line) and a
model with a heavy halo (dashed-dotted line, $M_{200} = 5.2 \times
10^{12} M_\odot$). These models respectively underpredict and
overpredict the observed velocity dispersion.}
\label{fig:fit_ngc3379}
\end{figure}

Our best fitting model for NGC~3379 requires a halo mass of $M_{200} =
1.0 \times 10^{12} M_\odot$, as compared to a total stellar mass of
$1.0 \times 10^{11} M_\odot$. In Figure~\ref{fig:fit_ngc3379} we show
the kinematics of our best-fit model compared to our data at large
radii and the long-slit kinematics on the major axis. The fit is
excellent, even for the higher order moments $h_3$ and $h_4$. A model
without dark halo clearly underpredicts the observed velocity
dispersion and $h_4$ outside 1 $R_e$. For comparison also a model with
a halo heavier than our best fit model is shown: here the velocity
dispersion is overestimated. In Figure~\ref{fig:chi2_3379} we show the
$\chi^2$ values of our best fitting model. We see that with 3-$\sigma$
confidence, $M_{200} = 1.0^{+0.4}_{-0.3} \times 10^{12} M_\odot$ under
the assumptions mentioned above, while a model without dark halo is strongly
excluded.

\begin{figure}
\psfig{figure=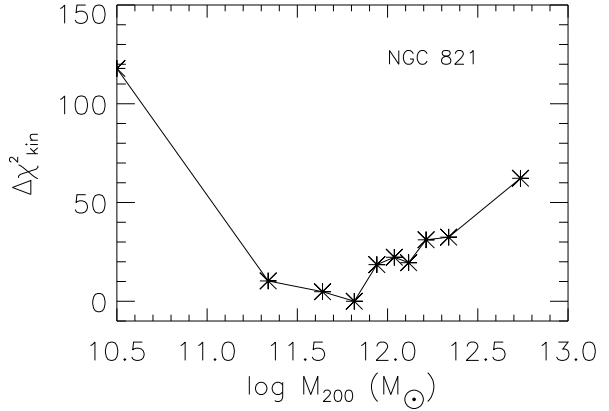,angle=0,width=8.0cm}
\caption{Same as Figure~\ref{fig:chi2_3379}, but now for NGC~821.}
\label{fig:chi2_821}
\end{figure}

\begin{figure}
\psfig{figure=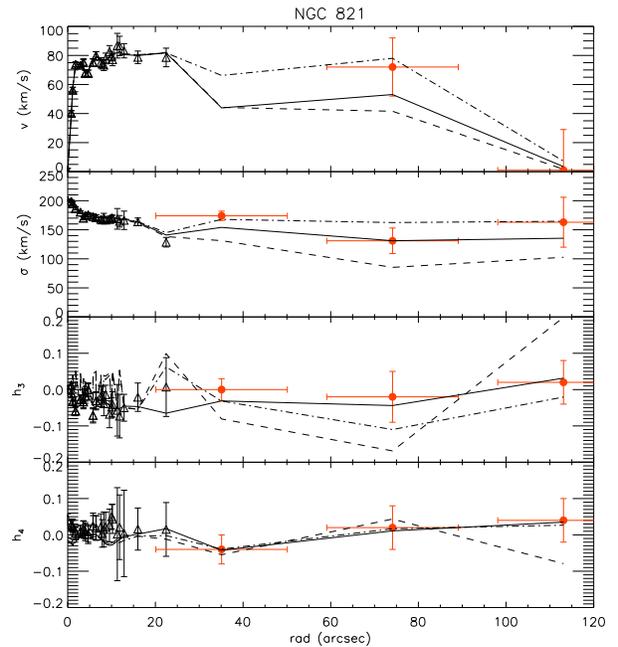,angle=0,width=8.0cm}
\caption{Best-fit model for NGC~821 (bold line, $M_{200} = 6.5 \times
10^{11} M_\odot$) compared to our datapoints at large radii (red
filled circles) and the central SAURON data (open triangles). The
dashed line indicates a model without halo and the dashed-dotted line
a model with a too heavy halo ($M_{200} = 5.5 \times 10^{12}
M_\odot$).}
\label{fig:fit_ngc821}
\end{figure}

We obtain similar results for NGC~821, where we also exclude the
absence of a dark halo (see Figure~\ref{fig:chi2_821}). The halo mass
in our best-fit model is $M_{200} = 6.5^{+1.0}_{-4.3} \times 10^{11}
M_\odot$, while the total stellar mass is $1.1 \times 10^{11}
M_\odot$. The kinematics of the fit compared to the observed
kinematics are shown in Figure~\ref{fig:fit_ngc821}, where for
comparison also a model without a dark halo and a model with a too
massive halo are shown.

We conclude that dark matter is present both in NGC~3379 and
NGC~821. Even within 1 $R_e$ in NGC~3379, 8 per cent of the total
mass is dark, while at 4 $R_e$ this fraction has increased to 34
per cent. NGC~821 is more dark matter dominated, with a dark matter
fraction of already 18 per cent within 1 $R_e$, increasing to 49 per
cent within 4 $R_e$. Note however that these values are based on
minimal halo assumptions, and therefore should be seen as lower
limits. Also, varying halo shape and profile can cause these
fractions to change, they therefore should be strictly seen as
indicative.

\subsection{Orbital structure}

\begin{figure*}
\begin{tabular}{cc}
\psfig{figure=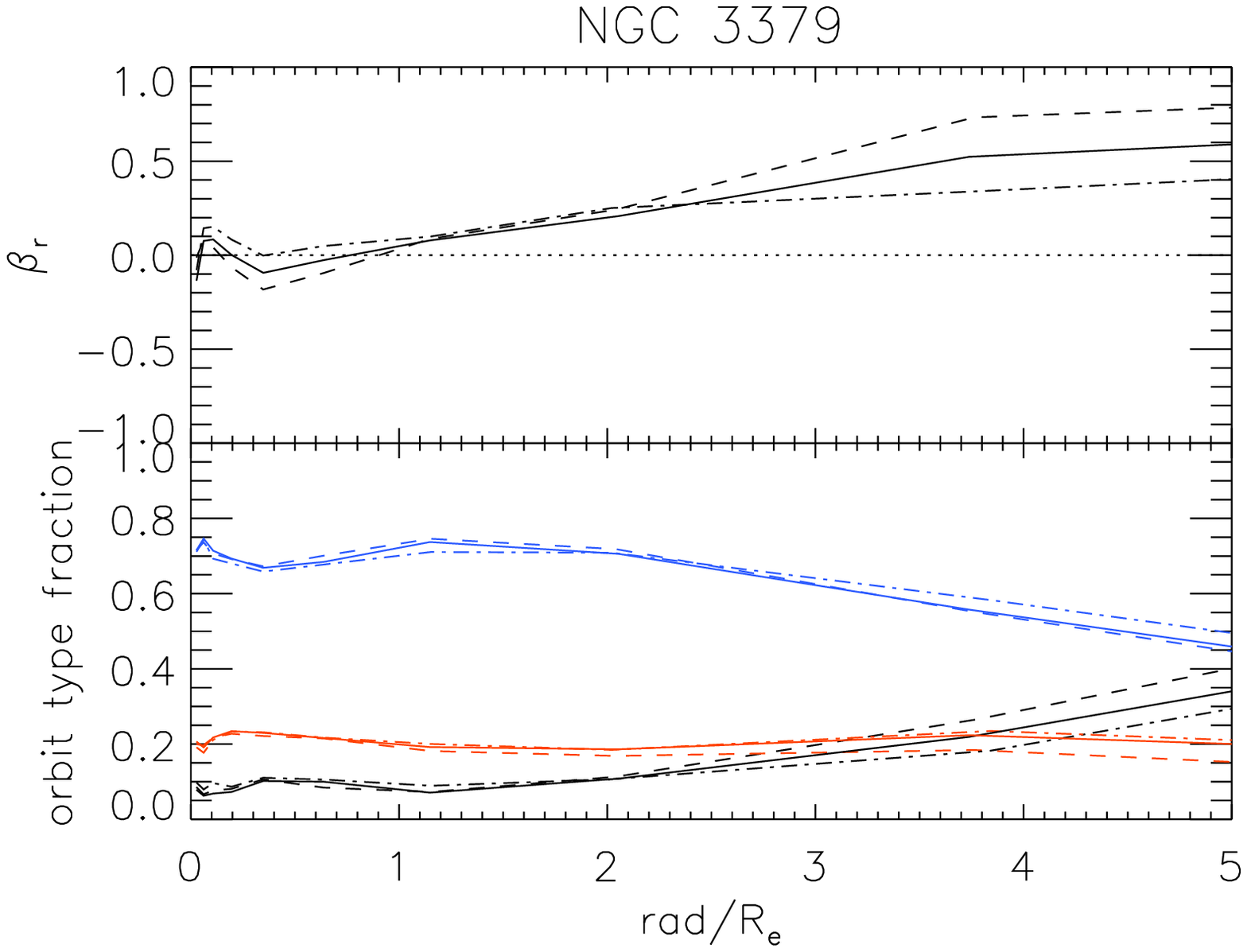,angle=0,width=7.5cm} &
\psfig{figure=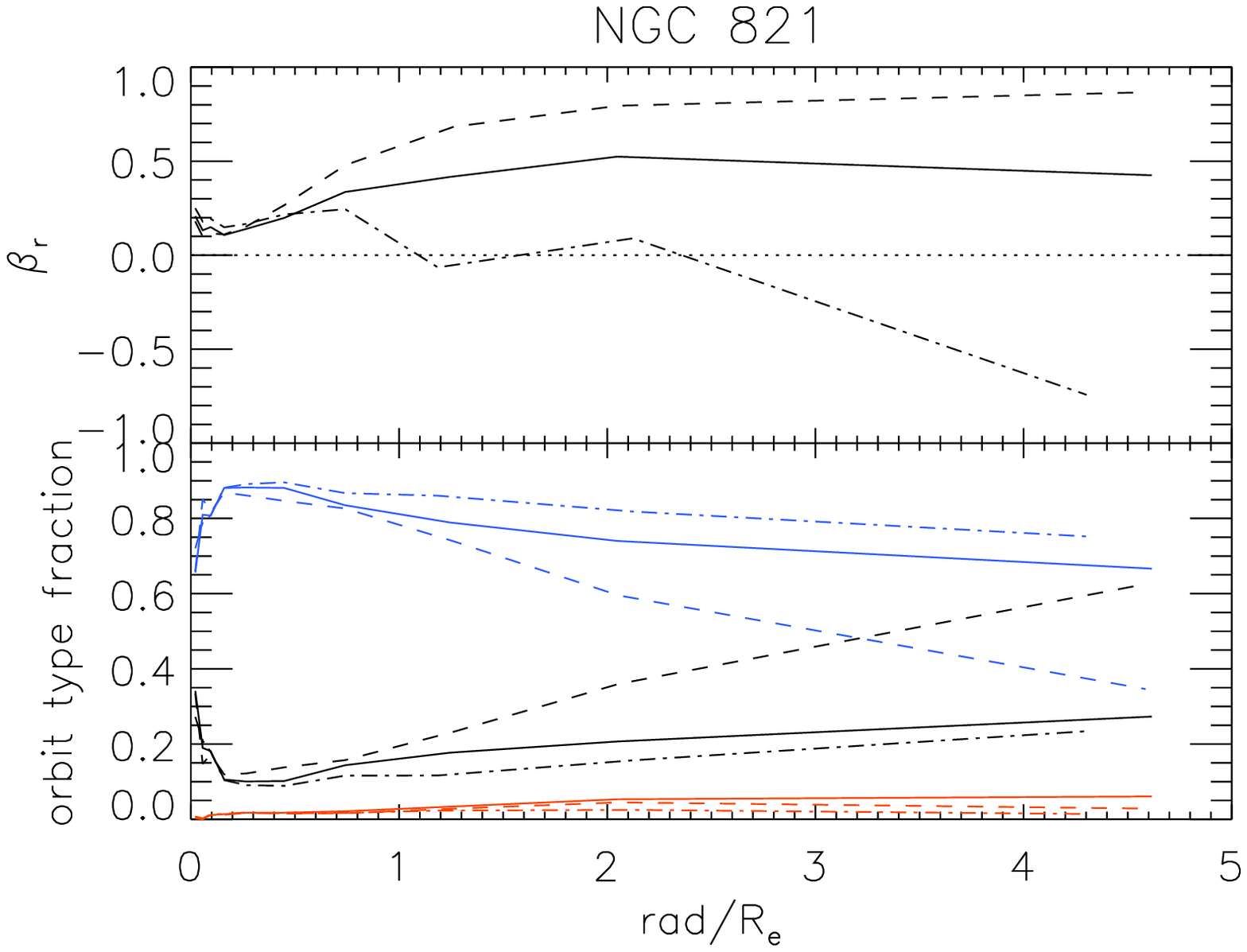,angle=0,width=7.5cm}
\end{tabular}

\caption{Orbital structure in NGC~3379 (left) and NGC~821 (right). Top
panel: velocity anisotropy $\beta_r$ as a function of radius (see
Equation~\ref{eq:betar}). The lines are as in
Figures~\ref{fig:fit_ngc3379} and \ref{fig:fit_ngc821}, with the solid
line for our best-fit model, the dashed line for the model without
halo and the dashed-dotted line for the model with the too heavy
halo. Bottom panel: fraction of orbit types as a function of radius
for the models of the top panel. Blue lines denote the fractions of
short axis tubes, red the fractions of long axis tubes and black the
fractions of box orbits.}
\label{fig:aniso}
\end{figure*}

In the top panels of Figure~\ref{fig:aniso} we show the orbital
anisotropy in our dynamical models of NGC~3379 and NGC~821. We
characterize the orbital anisotropy as

\begin{equation}
\beta_r = 1 - \frac{\sigma_\phi^2 + \sigma_\theta^2}{2\sigma_r^2},
\label{eq:betar}
\end{equation}

\noindent
with ($r,\theta, \phi$) standard spherical coordinates. Our best-fit
model of NGC~3379 is nearly isotropic within 2 $R_e$, and at larger
radii the radial anisotropy increases. Our best-fit model of NGC~821
is already mildly radially anisotropic in the central regions, but
also shows a small increase with radius.  Cappellari et
al. (2007)\nocite{2007MNRAS.379..418C} presented orbital anisotropy
values for both NGC~3379 and NGC~821 from axisymmetric models, based
on the same central SAURON data and MGE models as we use in this
paper. Our central anisotropy values are consistent with theirs for
both galaxies. Since Cappellari et al. (2007) used a different,
independent modeling code, this agreement gives confidence in the
robustness of the anisotropy parameters presented here.

Within 1 $R_e$, the orbital anisotropy changes only marginally between
our models with and without halo, consistent with the findings of
Thomas et al. (2009)\nocite{2009MNRAS.393..641T}. At larger radii the
radial anisotropy increases, and the increase is smaller for models
with heavier haloes. This is contrary to what is seen in spherical
models. In order to fit the observed mildly declining or constant
velocity dispersion profile without invoking extra dark matter at
large radii, the orbits have to be tangentially anisotropic, while
adding a dark halo results in more radially anisotropic orbits
(e.g. Richstone \& Tremaine 1984\nocite{1984ApJ...286...27R}; Dekel et
al. 2005\nocite{2005Natur.437..707D}; de Lorenzi et
al. 2009\nocite{2009MNRAS.395...76D}). 

The detailed kinematics in the inner part of the galaxy together with
our observed positive $h_4$ in the outer parts already tightly
constrain the fraction of box and tube orbits in our triaxial
potential, and therefore the anisotropy. Box orbits have no net
rotation, and contribute to the radial anisotropy of the system,
requiring $h_4 > 0$. For tube orbits to reproduce the observed nearly
zero mean rotation at large radii, they have to be counter-rotating,
increasing the velocity dispersion. However, they would at the same
time increase the tangential anisotropy and therefore yield $h_4 <
0$. Since we observe a positive $h_4$ at large radii, box orbits are
expected to contribute significantly to the orbital structure in our
models without dark matter, as confirmed in Figure~\ref{fig:aniso}. As
a result, the models without dark halo are strongly radial anisotropic
at large radii. Adding a spherical dark matter halo to the system
makes the gravitational potential less triaxial. This will decrease
the phase space volume of box orbits and naturally results in a larger
population of tube orbits (Hunter \& de Zeeuw
1992\nocite{1992ApJ...389...79H}). The radial anisotropy of the system
will decrease, as is illustrated in Figure~\ref{fig:aniso}. This
decrease is more pronounced in NGC~821 than in NGC~3379, since our
best-fit model of NGC~821 is more dark matter dominated than our
NGC~3379 model.

\subsection{\Mgb\ - \vesc\ relation}

We here investigate the connection between the stellar populations and
the mass distribution of the galaxy. Franx \& Illingworth
(1990)\nocite{1990ApJ...359L..41F} showed that there is a tight
relation between the local colour (tracing metallicity) and the local
escape velocity \vesc\ (tracing the gravitational potential $\Phi$) in
early-type galaxies, with \vesc\ given by:

\begin{equation}
\label{eq:vesc}
V_{\mathrm{esc}} = \sqrt{2\vert\Phi\vert}.
\end{equation}

\noindent
This relation was confirmed by Davies et
al. (1993)\nocite{1993MNRAS.262..650D}, who used \Mgtwee\ as a tracer
for metallicity. Scott et al. (2009)\nocite{2009MNRAS.submS} find
correlations between line strength indices and \vesc\ for the
early-type galaxies in the SAURON sample, based on measurements within
$\sim$1 $R_e$. They also show that for \Mgb\ this relation does not
only hold locally in one galaxy, but is also a global relation between
different galaxies.

\begin{figure}
\centerline{\psfig{figure=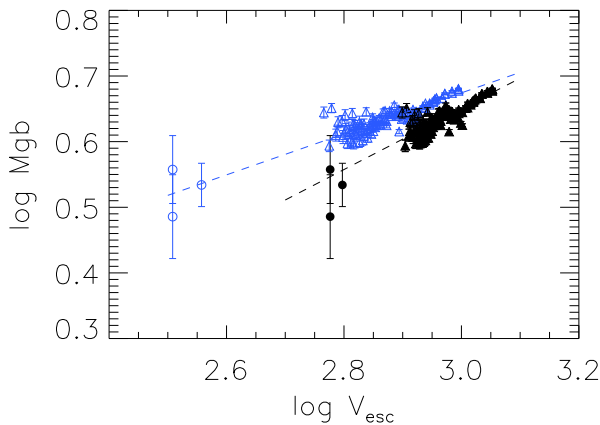,angle=0,width=8.0cm}}
\caption{\Mgb\ (in \AA) versus local escape velocity \vesc\ (km/s) in
NGC~3379. Black filled triangles denote points from the central SAURON
data, while the black filled dots are our large radii
data. Overplotted is the gradient fitted to the central dataset. With
open symbols (in blue) we show the \Mgb\ - \vesc\ relation calculated
from our model without a halo. This relation is offset from our
best-fit model and also has a different slope, illustrating the
influence of the dark matter.}
\label{fig:vesc}
\end{figure}

With our data at large radii we can now investigate the behaviour of
this relation out to 4 $R_e$. We determine for our best-fit model of
NGC~3379 the contribution to the gravitational potential of the dark
halo using Equation~\ref{eq:NFWpot}. The contribution of the stellar
mass is inferred from the MGE model, following the formularium in the
appendix of Cappellari et al. (2002)\nocite{2002ApJ...578..787C} and
the black hole is included as a Plummer potential. We plot our
result in Figure~\ref{fig:vesc}, and find that the observed relation
within 1 $R_e$ continues out to at least 4 $R_e$. The influence of the
dark halo on the \Mgb\ - \vesc\ relation becomes clear when comparing
our results to a model without a dark halo, where the only
contribution to the potential comes from the MGE model and the black
hole (open symbols in Figure~\ref{fig:vesc}). Not only is there an
offset in \vesc\ with respect to the model with the dark halo, but
also the slope changes, an effect which was also noted by Franx \&
Illingworth (1990)\nocite{1990ApJ...359L..41F}.  The addition of extra
mass at larger radii by the dark halo deepens the potential well,
which means that even particles in the centre of the galaxy will need
a larger \vesc\ to climb out of this well, causing a shift in \vesc\
at all radii. However, particles in the outskirts of the galaxy are
relatively more affected by the addition of the dark halo than
particles in the more central regions: if the halo had not been
present, escaping from the potential well at large radii would have
been relatively easy. Therefore, the shift in \vesc\ is larger at
larger radii, something that is also illustrated by
Figure~\ref{fig:vesc}. The net result is a change in slope. We find
that the slope of the \Mgb\ - \vesc\ relation in NGC~3379 changes in
logscale from 0.31 to 0.46, when taking the dark halo into account. If
instead of considering the infinite halo mass of the NFW profile we
only take the contribution of the halo within $r_{200}$ into account
when calculating the gravitational potential, the slope becomes 0.43
in logscale.

We note that in both the models with and without halo, there is a {\it
local} relation between \Mgb\ and \vesc. However, when looking at a
larger sample of galaxies and studying the {\it global} \Mgb\ - \vesc\
relation the contribution of the halo to the potential should be taken
into account to interpret the slopes, since the relative contribution
of the dark matter to the potential compared to the luminous one may
vary from galaxy to galaxy, e.g. depending on galaxy mass or
environment.

%%%%%%%%%%%%%%%%%%%%%%%%%%%%%%%%%%%%%%%%%%%%%%%%%%%%%%%%%%%%%%%%%%

\section{Summary and conclusion}
\label{sec:conclusion}

We used the SAURON IFU as a ``photon collector'' to obtain stellar
absorption line kinematics and line strengths out to four effective
radii in the early-type galaxies NGC~3379 and NGC~821. By co-adding
the individual spectra of each lenslet into one spectrum for each
observed field, we were able to obtain sufficient signal-to-noise in
the outskirts of these galaxies, which are too faint to observe with
traditional long-slit spectroscopy. The stellar velocity profiles are
measured up to the fourth Gauss-Hermite moment ($h_4$), which allows
us to break the mass-anisotropy degeneracy when constructing dynamical
models of our galaxies. The kinematics show a smooth continuation of
the profiles measured within 1-2 $R_e$ with SAURON and long-slit
spectroscopy, and are consistent with planetary nebulae
kinematics. The line strengths also show a continuation of the
gradients observed within 1 $R_e$, although our Fe5015 measurements
probably suffer from variations in the continuum shape. The stellar
halo population is old and metal-poor. The continuation of the line
strength gradients and the \Mgb\ - \vesc\ relation out to at least 4
$R_e$ provides constraints for the star formation and merger history
of early-type galaxies (e.g. Hopkins et
al. 2009a\nocite{2009ApJS..181..135H}; Hopkins et
al. 2009b\nocite{2009ApJS..181..486H}; Spolaor et
al. 2009\nocite{2009ApJ...691L.138S}).

Constructing dynamical orbit-based models, we show that even assuming
a maximal stellar contribution, both NGC~3379 and NGC~821 require a
dark halo to fit the observed kinematics. We parametrised the halo
contribution with a spherical NFW profile, fixing the concentration to
$c=10$. We therefore cannot give a definite number for the precise
mass content, shape and anisotropy of the halo. Varying $c$ in our
halo profile does not have a large effect on the fitted dark matter
content, as an increase of $c$ will be accompanied by a decrease in
$M_{200}$, as the dynamical models constrain the density inside the
region for which kinematics are available. Only with data at even
larger radii (preferably out to $\sim$ 100 $R_e$, accessible with
e.g. weak gravitational lensing or satellites) we would be able to
break the degeneracy between $c$ and $M_{200}$. In that sense, our
results should be viewed as an estimate of the minimal contribution of
the dark halo to the total mass budget in the galaxy, and we then see
that even at small radii (within 1 $R_e$), the dark matter is already
responsible for 10 - 20 per cent of the total mass. These percentages
will increase if we relax the minimal halo assumption, but we then
need reliable measurements of the stellar mass-to-light ratio and its
potential variation with radius. With better spatial coverage of the
galaxy, either by mosaicing or by employing larger IFUs (such as
VIRUS-P and in the future MUSE) and combining the so-obtained stellar
kinematics with other datasets tracing dark matter at different
regimes (e.g. \HI\ discs or rings, hot gas, gravitational lensing
and/or discrete tracers such as planetary nebulae) we will be able to
build more sophisticated dynamical models, to explore the shape and
mass of the dark haloes around early-type galaxies.

%%%%%%%%%%%%%%%%%%%%%%%%%%%%%%%%%%%%%%%%%%%%%%%%%%%%%%%%%%%%%%%%%%%%

\section*{Acknowledgements}

It is a pleasure to thank Eveline van Scherpenzeel, Chris Benn and the
ING staff for support on La Palma.  We gratefully acknowledge Koen
Kuijken, Aaron Romanowsky and Marc Sarzi for fruitful discussions. We
also thank Richard Wilman for supplying us with the data necessary to
make the superflat for NGC~3379 and Lodovico Coccato for sending us
the planetary nebulae kinematic profiles of NGC~821 and NGC~3379. We
are grateful to the referee, Elena Terlevich, for constructive
comments which improved the presentation of the paper.

This research was supported by the Netherlands Organization of
Scientific Research (NWO) through grants 614.000.426 and
614.000.301. AW acknowledges The Leids Kerkhoven-Bosscha Fonds for
contributing to working visits, as well as the hospitality of ESO, the
University of Oxford, l'Observatoire de Lyon and the Institute for
Advanced Study. MC acknowledges support from a PPARC Advanced
Fellowship (PP/D005574/1). GvdV acknowledges support through Hubble
Fellowship grant HST-HF-01202.01-A awarded by the Space Telescope
Science Institute, which is operated by the Association of
Universities for Research in Astronomy, Inc., for NASA, under contract
NAS 5-26555. 

The SAURON observations were obtained at the William Herschel
Telescope, operated by the Isaac Newton Group in the Spanish
Observatorio del Roque de los Muchachos of the Instituto de
Astrof\'isica de Canarias.

%%%%%%%%%%%%%%%%%%%%%%%%%%%%%%%%%%%%%%%%%%%%%%%%%%%%%%%%%%%%%%%%%%%%

%%%%%%%%%%%%%%%%%%%%%%%%%%%%%%%%%%%%%%%%%%%%%%%%%%%%%%%%%%%%%%%%%%%%%

\label{lastpage}

\begin{thebibliography}{}

\bibitem{2001MNRAS.326...23B} 
Bacon R. et al., 2001, MNRAS, 326, 23 

\bibitem{1998NewAR..42..503B}
Benn, C.R., Ellison, S.L., 1998, NewAr, 42, 503 

\bibitem{1993ApJ...416L..45B}
Bertola F., Pizzella A., Persic M., Salucci P., 1993, ApJ, 416L, 45

\bibitem{1982MNRAS.200..361B}
Binney J., Mamon G.A., 1982, MNRAS, 200, 361

\bibitem{2004MNRAS.351...63B}
Bower R.G., et al., 2004, MNRAS, 351, 63

\bibitem{2006MNRAS.373..157B}
Bridges T., et al., 2006, MNRAS, 373, 157

\bibitem{2001MNRAS.321..559B}
Bullock J.S., Kolatt T.S., Sigad Y., Sommerville R.S., Kravtsov A.V., Klypin A.A., Primack J.R., Dekel A., 2001, MNRAS, 321, 559 

\bibitem{1991ApJ...371..535C}
Capaccioli M., Vietri M., Held E.V., Lorenz H., ApJ, 1991, 371, 535

\bibitem{2002MNRAS.333..400C}
Cappellari M., 2002, MNRAS, 333, 400

\bibitem{2002ApJ...578..787C}
Cappellari M., Verolme E.K, van der Marel R.P., Verdoes Kleijn G.A., Illingworth G.D., Franx M., Carollo C.M., de Zeeuw P.T., 2002, ApJ, 371, 535

\bibitem{2004PASP..116..138C}
Cappellari M., Emsellem E., 2004, PASP, 116, 138

\bibitem{2006MNRAS.366.1126C}
Cappellari M., et al., 2006, MNRAS, 366, 1126

\bibitem{2007MNRAS.379..418C}
Cappellari M., et al., 2007, MNRAS, 379, 418

\bibitem{1984ApJ...286..403C}
Carlberg R.G., 1984, ApJ, 286, 403

\bibitem{1995ApJ...441L..25C}
Carollo C.M., de Zeeuw P.T., van der Marel R.P., Danziger I.J., Qian E.E., 1995, ApJ, 441, L25

\bibitem{2009MNRAS.394.1249C}
Coccato L., et al., 2009, MNRAS, 394, 1249

\bibitem{2003ApJ...591..850C}
C\^ot\'e P., McLaughlin D.E., Cohen J.G., Blakeslee J.P., 2003, ApJ, 591, 850

\bibitem{1993MNRAS.262..650D}
Davies R.L., Sadler E.M., Peletier R.F., 1993, MNRAS, 262, 650

\bibitem{2005Natur.437..707D}
Dekel A., Stoehr F., Mamon G.A., Cox T.J., Novak G.S., Primack J.R., 2005, Nature, 437, 707 

\bibitem{2009MNRAS.395...76D}
de Lorenzi F., et al., 2009, MNRAS, 395, 76

\bibitem{2002MNRAS.329..513D}
de Zeeuw P.T., et al., 2002, MNRAS, 329, 513


\bibitem{2007ApJ...664..257D}
Douglas N.G., et al., 2007, ApJ, 664, 257

\bibitem{1994A&A...285..723E}
Emsellem E., Monnet G., Bacon R., 1994, A\&A, 285, 723

\bibitem{2004MNRAS.352..721E}
Emsellem E., et al., 2004, MNRAS, 352, 721 

\bibitem{2007MNRAS.379..401E}
Emsellem E., et al., 2007, MNRAS, 379, 401

\bibitem{2008arXiv0803.3626F}
Forestell A., Gebhardt K, 2008, arXiv0803.3626

\bibitem{1990ApJ...359L..41F}
Franx M, Illingworth G., 1990, ApJ, 359L, 41

\bibitem{1994ApJ...436..642F} 
Franx M., van Gorkom J.H., de Zeeuw P.T, 1994, ApJ, 436, 642

\bibitem{2001ApJ...553...47F}
Freedman W.L., et al., 2001, ApJ, 553, 47

\bibitem{2007ApJ...667..176G}
Gavazzi R., Treu T., Rhodes J.D., Koopmans L.V.E., Bolton A.S., Burles S., Massey R.J., Moustakas L.A., 2007, ApJ, 436, 176

\bibitem{1993MNRAS.265..213G}
Gerhard O.E., 1993, MNRAS, 265, 231

\bibitem{2001AJ....121.1936G}
Gerhard O., Kronawitter A., Saglia R.P., Bender, R., 2001, AJ, 121, 1936 

\bibitem{2009ApJS..181..135H}
Hopkins P.F., Cox T.J., Dutta S.N., Hernquist L., Kormendy J., Lauer T.R., 2009a, ApJS, 181, 135

\bibitem{2009ApJS..181..486H}
Hopkins P.F., Lauer T.R. Cox T.J., Hernquist L., Kormendy J., 2009b, ApJS, 181, 486

\bibitem{2006ApJ...646..899H}
Humphrey P.J., Buote D.A., Gastaldello F., Zappacosta L., Bullock J.S., Brighenti F., Mathews W.G., 2006, ApJ, 646, 899 

\bibitem{1992ApJ...389...79H}
Hunter C., de Zeeuw P.T., 1992, ApJ, 389, 79

\bibitem{2002SciAm.286f..36K}
Kauffmann G., van den Bosch F., 2002, SciAm, 286, 36

\bibitem{2006ApJ...649..599K}
Koopmans L.V.E., Treu T., Bolton A.S., Burles S., Moustakas L.A., 2006, ApJ, 649, 599

\bibitem{2008MNRAS.390...93K}
Krajnovi\'c D., et al., 2008, MNRAS, 390, 93

\bibitem{2006MNRAS.369..497K}
Kuntschner H., et al., 2006, MNRAS, 369, 497

\bibitem{2005MNRAS.362..799M}
Maraston C., 2005, MNRAS, 362, 799

\bibitem{2002PhD}
McDermid R.M., 2002, PhD thesis, University of Durham 

\bibitem{2005ApJ...625..121M}
Mei S., et al., 2005, ApJ, 625, 121

\bibitem{2006MNRAS.371..157M}
Morganti, R. et al., 2006, MNRAS, 371, 157

\bibitem{1996ApJ...462..563N}
Navarro J.F., Frenk C.S., White S.D., 1996, ApJ, 462, 563

\bibitem{2009MNRAS.393..329N}
Napolitano N.R., et al., 2009, MNRAS, 393, 329

\bibitem{2002AJ....123..729O}
Oosterloo T.A., Morganti R., Sadler E.M., Vergani D., Caldwell N., 2002, AJ, 371, 729

\bibitem{2004MNRAS.354..935O}
O'Sullivan E., Ponman T.J., 2004, MNRAS, 354, 935

\bibitem{2005MNRAS.362..857P}
Proctor R.N., Forbes D.A., Forestell A., Gebhardt K., 2005, MNRAS, 362, 857

\bibitem{1984ApJ...286...27R}
Richstone D.O., Tremaine S., 1984, ApJ, 286, 27

\bibitem{1997ApJ...488..702R}
Rix H.-W., de Zeeuw P.T., Cretton N., van der Marel R.P., Carollo C.M., 1997, ApJ, 488, 702 

\bibitem{2003Sci...301.1696R}
Romanowsky A.J., Douglas N.G., Arnaboldi M., Kuijken K., Merrifield M.R., Napolitano N.R., Capaccioli M., Freeman K.C., 2003, Science, 301, 1696  

\bibitem{2006MNRAS.371..703S}
S\'anchez-Bl\'azquez P., et al., 2006, MNRAS, 371, 703

\bibitem{2007MNRAS.377..759S}
S\'anchez-Bl\'azquez P., Forbes D.A., Strader J., Brodie J., Proctor R., 2007, MNRAS, 377, 759

\bibitem{2006MNRAS.366.1151S}
Sarzi M., et al., 2006, MNRAS, 366, 1151

\bibitem{2007ApJS..171..146S}
Schiavon R.P., 2007, ApJS, 171, 146

\bibitem{2009MNRAS.submS}
Scott N., et al., 2009, MNRAS, submitted

\bibitem{2006MNRAS.370..559S}
Shapiro K.L., Cappellari M., de Zeeuw P.T., McDermid R.M., Gebhardt K., van den Bosch R.C.E., Statler T.S., 2006, MNRAS, 370, 559

\bibitem{2009ApJ...691L.138S} 
Spolaor M., Proctor R.N., Forbes D.A., Couch W.A., 2009, ApJ, 691L, 138

\bibitem{2005Natur.435..629S}
Springel V., et al., 2005, Nature, 435, 629

\bibitem{1999AJ....117..839S}
Statler T.S., Smecker-Hane T., 1999, AJ, 117, 839

\bibitem{2001AJ....121..244S}
Statler T.S., 2001, AJ, 121, 244

\bibitem{2003MNRAS.339..897T}
Thomas D., Maraston C., Bender R., 2003, MNRAS, 339, 897

\bibitem{2007MNRAS.382..657T}
Thomas J., Saglia R.P., Bender R., Thomas D., Gebhardt K., Magorrian J., Corsini E.M., Wegner G., 2007, MNRAS, 382, 657

\bibitem{2009MNRAS.393..641T}
Thomas J., et al., 2009, MNRAS, 393, 641

\bibitem{2001ApJ...546..681T}
Tonry J.L., Dressler A., Blakeslee J.P., Ajhar E.A., Fletcher A.B., Luppino G.A, Metzger M.R., Moore C.B., 2001, ApJ, 546, 681

\bibitem{1985ApJ...295..305V}
van Albada T.S., Bahcall J.N, Begeman K., Sancisi R., 1985, ApJ, 295, 305

\bibitem{2008MNRAS.385..614V}
van de Ven G., de Zeeuw P.T., van den Bosch R.C.E., 2008, MNRAS, 385, 614

\bibitem{2008MNRAS.385..647V}
van den Bosch R.C.E., van de Ven G., Verolme E.K., Cappellari M., de Zeeuw P.T., 2008, MNRAS, 385, 647

\bibitem{2008arXiv0811.3474V}
van den Bosch R.C.E., van de Ven G., 2008, arXiv0811.3474


\bibitem{2009MNRAS.subm}
van den Bosch R.C.E., de Zeeuw P.T., 2009, MNRAS, submitted

\bibitem{2007IAUS..241..133V}
Vazdekis A., et al., 2007, in Vazdekis A., Peletier R.F., ed., Proc. IAU Symp. 241, Stellar Populations as Building Blocks of Galaxies. Cambridge University Press, Cambridge, p. 133

\bibitem{1980MNRAS.191P...1W}
White S.D.M., 1980, MNRAS, 191, 1p

\bibitem{2008MNRAS.383.1343W}
Weijmans A., Krajnovi\'c D., van de Ven G., Oosterloo T.A., Morganti R., de Zeeuw P.T., 2008, MNRAS, 383, 1343 




\end{thebibliography}
\end{document}